\documentclass[]{aastex631}

\usepackage{multirow} 
\usepackage{amsmath}
\usepackage{booktabs}

\begin{document}

\title{Constraining the Nuclear Equation of State of a Neutron Star via High-frequency Quasi-periodic Oscillation in Short Gamma-Ray Bursts}
    
    \author{Jun-Xiang Huang}
        \affiliation{Guangxi Key Laboratory for Relativistic Astrophysics, School of Physical Science and Technology, Guangxi University, Nanning 530004, People’s Republic of China}
        
   \author{Hou-Jun L\"{u}}
   \altaffiliation{Corresponding author email: lhj@gxu.edu.cn}
   \affiliation{Guangxi Key Laboratory for Relativistic Astrophysics, School of Physical Science and Technology, Guangxi University, Nanning 530004, People’s Republic of China}
   
    \author{En-Wei Liang}
        \affiliation{Guangxi Key Laboratory for Relativistic Astrophysics, School of Physical Science and Technology, Guangxi University, Nanning 530004, People’s Republic of China}


    \date{\today}

\begin{abstract}
The determination of the equation of state (EOS) of a neutron star (NS) and its maximum mass is very important for understanding the formation and properties of NSs under extreme conditions, but they remain open questions. Short-duration gamma-ray bursts (GRBs) are believed to originate from the merger of binary NSs or giant flares (GFs) of soft gamma repeaters (SGRs). Recently, the high-frequency quasi-periodic oscillations (QPOs) have been claimed to be identified from two short GRBs (GRB 931101B and GRB 910711). In this paper, we propose that the observed high-frequency QPOs in these two short GRBs result from torsional oscillations in the GFs of SGRs associated with cold NSs, or from radial oscillations of hypermassive NSs as the hot remnants of binary NS mergers, and then to constrain the EOS of NSs. For torsional oscillations, the six selected EOSs (TM1, NL3, APR, SLy4, DDME2, and GM1) of NSs suitable for the zero-temperature condition exhibit significant overlap in mass ranges, suggesting that we cannot constrain the EOS of NSs. For radial oscillations, the six selected EOSs (IUF, TM1, TMA, FSG, BHBLp, and NL3) of NSs suitable for the high-temperature condition cannot be ruled out when redshift is considered. However, it is found that the EOS can only be constrained if the redshift and temperature of the remnant can be measured.
\end{abstract}

\section{Introduction} \label{sec:intro}
    Neutron stars (NSs) are one of the most intriguing compact objects in astrophysical studies, due to their extreme properties. The maximum mass and equation of state (EOS) of NSs play an important role in understanding their structure and properties \citep{Shapiro.et.al.1983}. The first estimate of the maximum mass of an NS ($M_{\text{max}}$) for a static equilibrium solution in early studies was $M_{\text{max}}\sim\, 0.72\ M_\odot$ Me by simplifying the NS as a cloud of noninteracting cold Fermi gas \citep{Oppenheimer.Volkoff.1939}. However, from an observational point of view, subsequent precise measurements of NS mass from binary systems have revealed maximum masses that exceed the theoretical predictions \citep{Hulse.Taylor.1975, Anderson.et.al.1990, Wolszczan.1991, Burgay.et.al.2003}, and this suggests that the interactions between particles in the core cannot be ignored. Due to the poorly known and uncertain composition and interactions in the core of NSs, as well as the EOS of NSs, the maximum mass $M_{\text{max}}$ of an NS remains highly uncertain, ranging from $1.46\,M_\odot$ to $2.48\,M_\odot$ for different EOSs \citep{Haensel.et.al.2007}. So far, the maximum mass of an NS that we have accurately observed is ${2.08}_{-0.07}^{+0.07}\ {M}_{\odot }$ (68.3\% credibility) from PSR J0740+6620, determined by relativistic Shapiro delay \citep{Fonseca.et.al.2021}.
     
    Several previous studies have attempted to constrain the EOS of NSs, such as through the observed internal plateau together with the collapse time in the X-ray emission of short-duration gamma-ray bursts (GRBs; \citet{Lasky.et.al.2014, L"u.at.al.2015}) , gravitational-wave (GW) radiation from hypermassive NSs (HMNS; \citet{Shibata.2005}), GW radiation from supramassive NSs \citep{Lan.et.al.2020}, GW radiation from the merger of NS binaries \citep{Takami.et.al.2014}, and the threshold mass for prompt black  hole (BH) formation from double NS mergers \citep{Bauswein.et.al.2020}. However, the constraints from these results are not yet sufficient. Thus, understanding the EOS of NSs remains an open question in astrophysics and particle physics \citep{Lattimer.2021}.

    On the other hand, by considering the NS as a good resonator, numerous quasi-periodic oscillation (QPOs) modes of the NS would be excited. Based on their restoring forces, the oscillation modes can be categorized into radial and different nonradial modes (such as fundamental mode, pressure mode, gravity mode, Rossby mode, and shear mode; \citet{McDermott.et.al.1988,Stergioulas.2003}). \citet{Watts.Strohmayer.2007} discovered QPOs with frequencies ranging from tens of Hz up to a few kHz by analyzing the X-ray light curves of SGR 0526–66, SGR 1900+14, and SGR 1806–20. The discovery of QPOs in the three cases is believed to be the first direct detection of magnetar oscillations, and it is naturally interpreted as crustal shear modes triggered by starquakes, which are associated with giant flares (GFs; \citet{Duncan.1998, Piro.2005, Watts.Strohmayer.2007, Lee.2007, Samuelsson.Andersson.2007, Sotani.et.al.2007,Sotani.et.al.2012,Steiner.et.al.2009}). The oscillations are strongly dependent on the density distribution of the stellar interior and the shear modulus within the elastic region. By comparing the observed frequencies with theoretical models, precise constraints can be placed on the properties of ultradense matter that cannot be replicated to check in terrestrial laboratories \citep{Lattimer.Prakash.2001,Sotani.et.al.2012}.
    
    Recently, \citet{Chirenti.et.al.2023} claimed that kHz quasi-periodic signals were identified in two short-duration GRBs (GRB 931101B and GRB 910711) from BATSE data (see Table \ref{tab1}). They proposed that the physical origin of these two GRBs may originate from the merger of binary NSs, and such a merger results in a hot remnant, such as an HMNS. If the QPOs of those two short-duration GRBs are indeed from the quasi-periodic modulation of gamma rays caused by the global oscillations, then it may provide a new opportunity to explore the properties of NSs by employing seismology. In this paper, we consider two potential mechanisms that may produce the observed high-frequency QPOs in two short-duration GRBs. One is torsional oscillation with crustal shear modes in cold NSs (Section 2), and the other one is radial oscillation in the hot merger remnants by considering the effect of temperature (Section 3). Then, we constrain the EOS of neutron stars (NSs) by adopting the observed high-frequency QPOs in those two short-duration GRBs (Section 3.2). Conclusions are drawn in Section 4, with some additional discussions.
        
 \section{Torsional Oscillation}
     Soft gamma repeaters (SGRs) are young, slowly spinning magnetars, and strong flaring activities exhibiting durations from milliseconds to seconds with QPOs can be produced during the active outburst phase \citep{Duncan.Thompson.1992}. Because of the intense $\gamma$-ray luminosity and the spectral characteristics, the emission of GFs and that of short-duration GRBs are quite similar, and GFs of magnetars were also detected as disguised short-duration GRBs from nearby galaxies. Recently, two short GRBs (GRB 200415A and GRB 051103A) have been interpreted as giant magnetar flares \citep{Yang.et.al.2020, Svinkin.et.al.2021, Roberts.et.al.2021}. If this is the case, the GFs with QPOs are driven by elastic shear modes in the crust of a cold NS that is believed to consist of a fluid core with a radius $r_{\text{c}}$, and an elastic crust with a thickness of $\sim$ 1 km accounts for a mass of $\sim\,0.01\ {M}_{\odot}$. As strong magnetic fields and/or rotational effects evolve with time, the stresses will build up in the crust. When these stresses reach a critical threshold, the crust will eventually fracture to trigger oscillations. Because the velocity field of torsional oscillations (i.e., the restoring force supported by shear force) is divergence-free with no radial component, it is not related to the varying density and pressure of the star \citep{Schumaker.Thorne.1983}. Thus, it is more easily excited with a fundamental frequency of $\sim$ 30 Hz after a crustquake of a magnetar with a strong magnetic field \citep{Schumaker.Thorne.1983,McDermott.et.al.1988,Duncan.1998}.
     
     One needs to note that \cite{Chirenti.et.al.2023} excluded an SGR origin, primarily based on energetics estimates from the redshift $z$ inferred using distances to galaxies with moderately active star formation, which are consistent with the burst localizations. However, this scenario cannot be entirely ruled out as a possibility in a low-redshift context. In this section, we hypothesize that the observed short-duration GRBs with high frequency QPOs (GRB 931103B and GRB 910711) are caused by GFs of magnetars at low redshift. Specifically, we consider $z < 0.015$, corresponding to a maximum isotropic energy $E_\text{iso} = 5.3 \times 10^{46}$ erg in the GF sample (see below for details). Within this assumption, we discuss the torsional oscillations of NSs and adopt the high-frequency QPOs observed in these bursts to constrain the EOS of NSs.
  
     \subsection{Physical Model}
         \citet {Thorne.Campolattaro.1967} first derived the equations governing the nonradial oscillation of NSs in general relativity. Similarly, we consider a spherical NS under the hypothesis of a nonrotating and nonmagnetic field. The marked feature of the torsional oscillation mode is that it does not involve radial displacement or induce any change in internal density and pressure. Consequently, within the frame of spherical coordinates ($r$, $\theta$ and $\phi$), the defined crystalline matter element moves only along the angle $\phi$. The magnitude of proper displacement in the $\phi$-direction of a material element during oscillation can be denoted as $\mathrm{d} l$, which can be expressed as
        \begin{equation}
            \mathrm{d} l=r Y(r) \exp (\mathrm{i} \omega t) b_{\ell}(\theta), \quad b_{\ell}(\theta)=\frac{\partial}{\partial \theta} P_{\ell}(\cos \theta) .
        \end{equation}
        Here $b_{\ell}(\theta)$ is related to the Legendre polynomial $P_{\ell}(\cos \theta)$, and it is the only $\theta$-dependent function that governs the angular behavior of displacement. Hence, $\ell$ is the angular index that determines the angular nodes of the oscillation. The comprehensive mathematical framework for torsional oscillations of nonrotating stars in the context of general relativity was established by \citet{Schumaker.Thorne.1983}. The function $Y(r)$ governs the radial behavior of the oscillation and can be formulated as \citep{Schumaker.Thorne.1983,Sotani.et.al.2012}
        \begin{equation}
            \label{eq:Y}
            Y^{\prime \prime}+\left(\frac{4}{r}+\Phi^{\prime}-\Lambda^{\prime}+\frac{\mu^{\prime}}{\mu}\right) Y^{\prime} 
            +\left[\frac{\rho+P / c^{2}}{\mu} \omega^{2} \mathrm{e}^{-2 \Phi}-\frac{(\ell+2)(\ell-1)}{r^{2}}\right] \mathrm{e}^{2 \Lambda} Y=0 .
        \end{equation}
        Here, $\Phi$ and $\Lambda$ are radial functions of the Schwarzschild metric 
        \begin{equation}
            \mathrm{d}s^{2}=-\mathrm{e}^{2 \Phi} \mathrm{d} t^{2}+\mathrm{e}^{2 \Lambda} \mathrm{d} r^{2}+r^2\mathrm{~d} \theta^{2}+r^2\sin ^{2} \theta \mathrm{d} \phi^{2}.
        \end{equation}
        Neglecting space-time perturbations, $\Phi$ and $\Lambda$ can be mathematically expressed as 
        \begin{equation}
            \exp \Phi(r)=\exp (-\Lambda(r))=-\frac{1}{\sqrt{1-2 G m(r) /\left(r c^{2}\right)}}.
        \end{equation}

        The shear modulus, denoted as $\mu$, plays a crucial role in the oscillations of the crust by acting as the necessary restorative force. The crust is treated as an isotropic body-centered cubic Coulomb solid, and a detailed calculation of the shear modulus is presented in \citet{Ogata.Ichimaru.1990}. For the cold-catalyzed crust with temperature $T=0$, $\mu$ is given as
        \begin{equation}
            \label{eq:mu}
            \mu=0.1194 \frac{n_{i}Z^{2} e^{2}}{a},
        \end{equation}
        where $n_{i}$ is the total number density of ions in the plasma matter, $a = (3 \pi n_{i}/4)^{1/3}$ is the average ion spacing, $Z$ is the charge number of a single ion, and $e$ is the elementary charge. The equilibrium nuclides $(A, Z)$ present in the cold-catalyzed matter at different densities are listed in \citet{Haensel.Pichon.1994} and \citet{ Negele.Vautherin.1973}. The discontinuity of $\mu$ occurs at the point where a change in nuclide composition takes place. To facilitate numerical calculations, we adopt the smooth composition model to simulate the shear modulus \citep{Douchin.Haensel.2000}.

        Finally, by giving the mass and EOS of an NS, the spherically symmetric density $\rho$ and pressure $P$ of the NS can be solved by the Tolman–Oppenheimer–Volkoff equation. Then, the eigenfrequency $\omega$ can be obtained by solving Equation~(\ref{eq:Y}), supplemented by the boundary conditions $Y^{\prime}(r_{\text{c}})=0$ and $Y^{\prime}(R)=0$ at the core–crust interface and the surface of the NS, respectively.

        In general, the oscillation of an NS has an infinite number of discrete eigenfrequency solutions. The different modes of torsional oscillation can be labeled as $_\ell t_n$, which correspond to the frequency $_\ell\omega_n=2\pi_\ell\nu_n$. Here the indices $n=0,1,2,...$ and $\ell=2,3,4,\ldots$ represent the nodes of oscillation in radial and angular directions, respectively. The specific excitation of oscillation modes is related to the properties of the crust and the processes of coupling and damping. However, some oscillation modes may be excited but not detected. The reason may be related to the modulation modes of the gamma-ray light curve, which remain unclear.
        
    \subsection{Constraint on the EOS of NSs}
        For the fundamental oscillations without radial nodes ($n = 0$), the momentum transfer of shear perturbations in the $\theta$-direction is significantly greater than that in the $r$-direction. If this is the case, it will result in the formation of standing waves during the propagation (with a typical length scale of $\sim \pi R$) of shear perturbations along the $\theta$-direction. In contrast, the length scale of propagation in the $r$-direction is $\sim\,0.1R$ \citep{Gabler.et.al.2012}. Consequently, this process takes a longer time to result in lower oscillation frequencies $_\ell\nu_0 \sim 20 - 100$ Hz, which are much lower than those of observed QPOs in the two shortduration GRBs GRB 931103B and GRB 910711 \citep{Chirenti.et.al.2023}. Hence, it is believed that the origin of QPO frequencies observed in those two short-duration GRBs is not likely to relate to fundamental oscillations. On the other hand, by considering the eigenfrequency from ordinary oscillations with $n\ge1$ (one or more radial nodes) as a function of mass, one can obtain the “fine splitting” structure, which is similar to a spectrum. The frequencies change slightly with increasing values of $\ell$ compared to the change with $n$. Therefore, we adopt $\ell = 2$ to replace the other modes with $\ell > 2$. Initially, we consider only cold-catalyzed (zero-temperature), nonmagnetic, nonrotating spherically symmetric models. The given mass of an NS is strongly dependent on the frequency spectrum. In this case, we adopt a 0.5\% error in the frequency spectrum lines, which may be caused by the greater complexity of the actual NS (e.g., temperature, magnetic fields, rotation, and nonspherically symmetric features), but our knowledge about these complexities is limited.
        
\begin{figure*}
            \centering
            \includegraphics[width=0.32\textwidth]{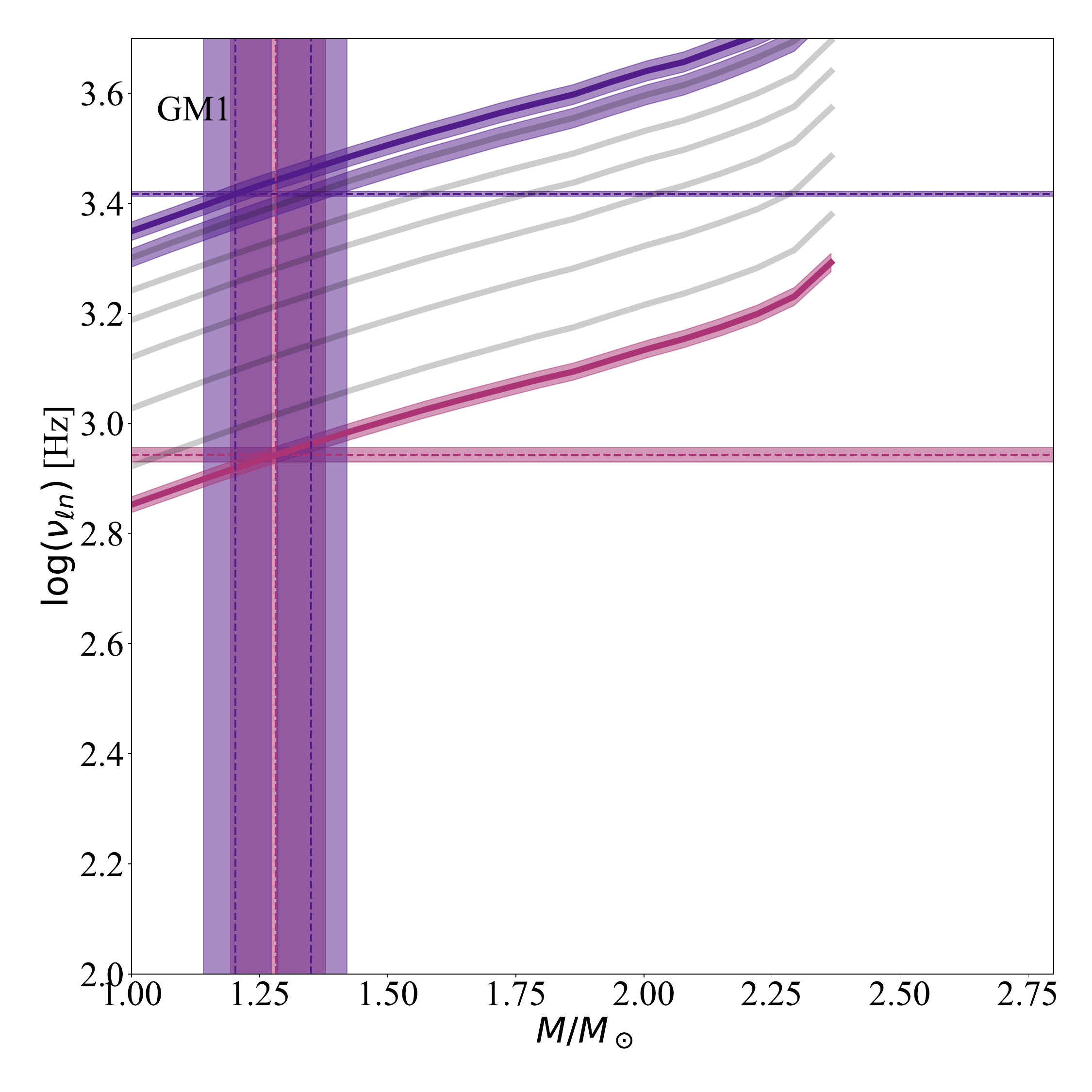}
            \includegraphics[width=0.32\textwidth]{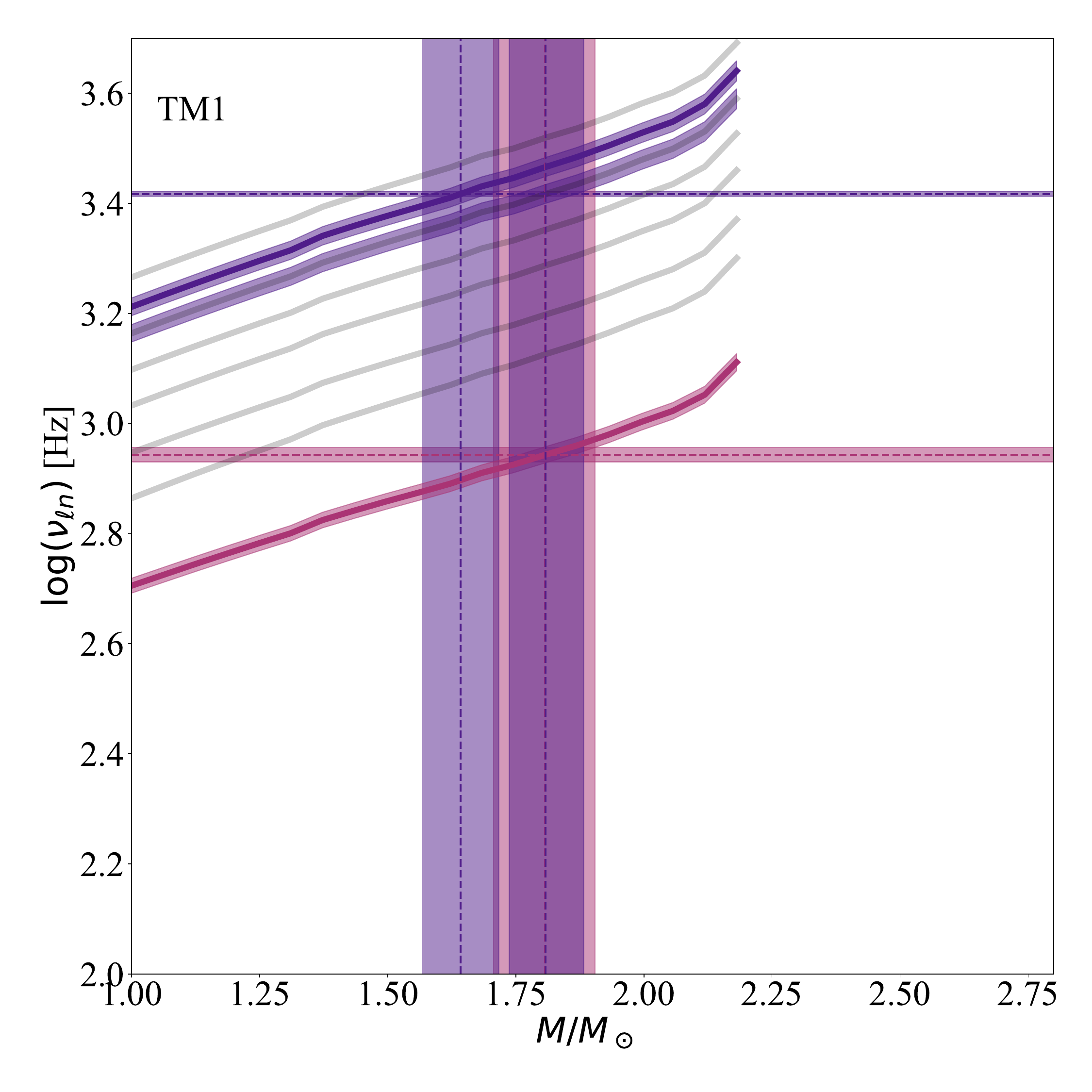}
            \includegraphics[width=0.32\textwidth]{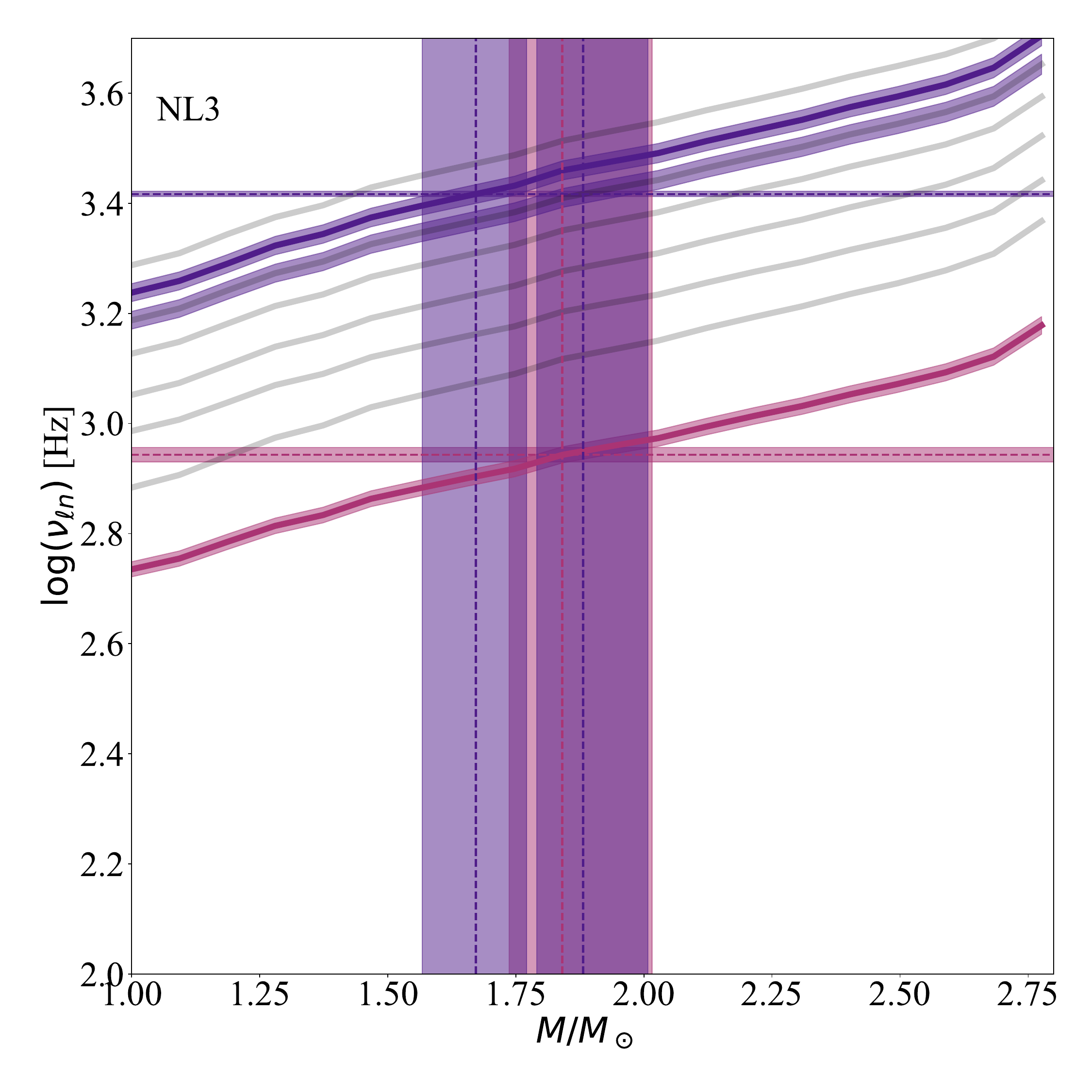}\\
            \includegraphics[width=0.32\textwidth]{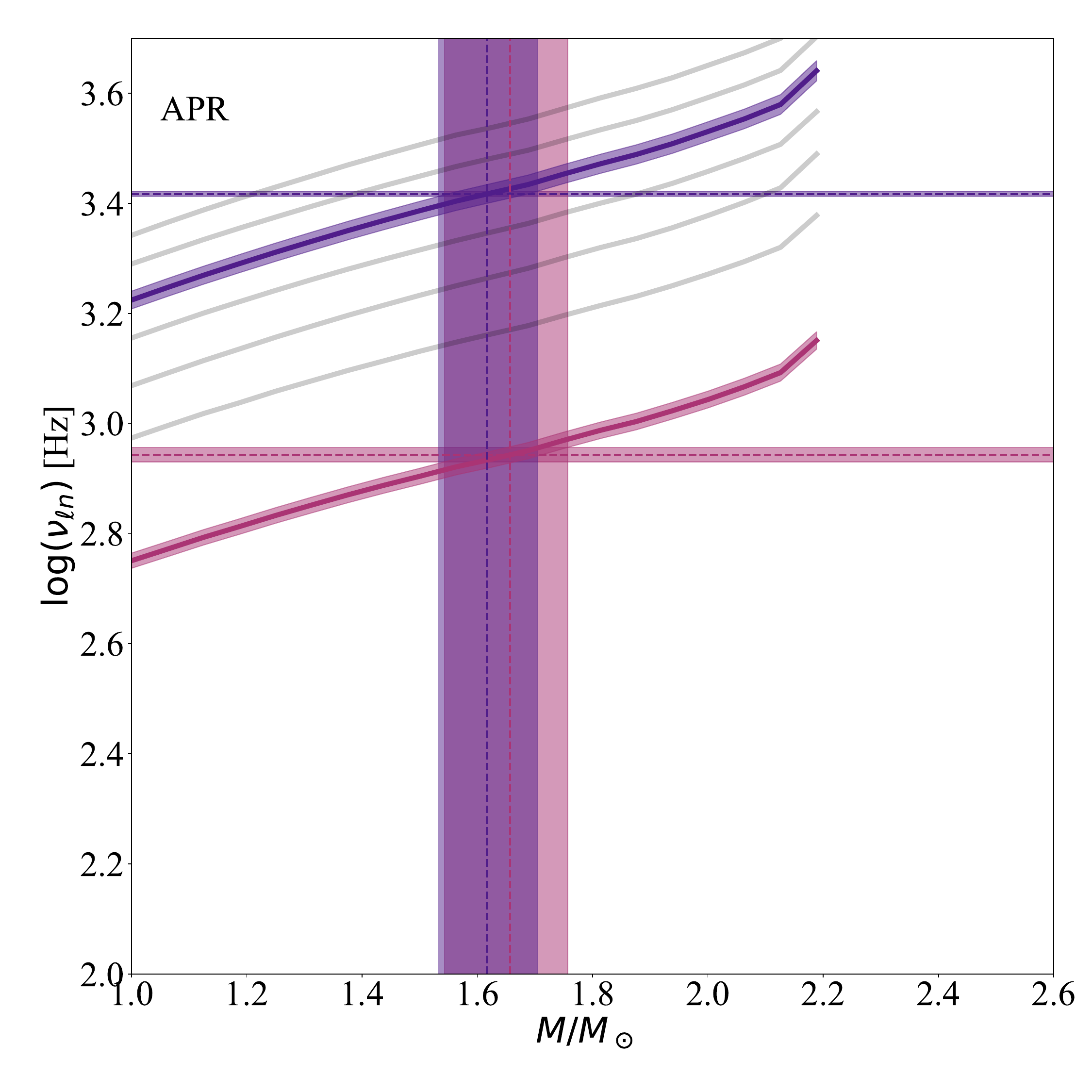}
            \includegraphics[width=0.32\textwidth]{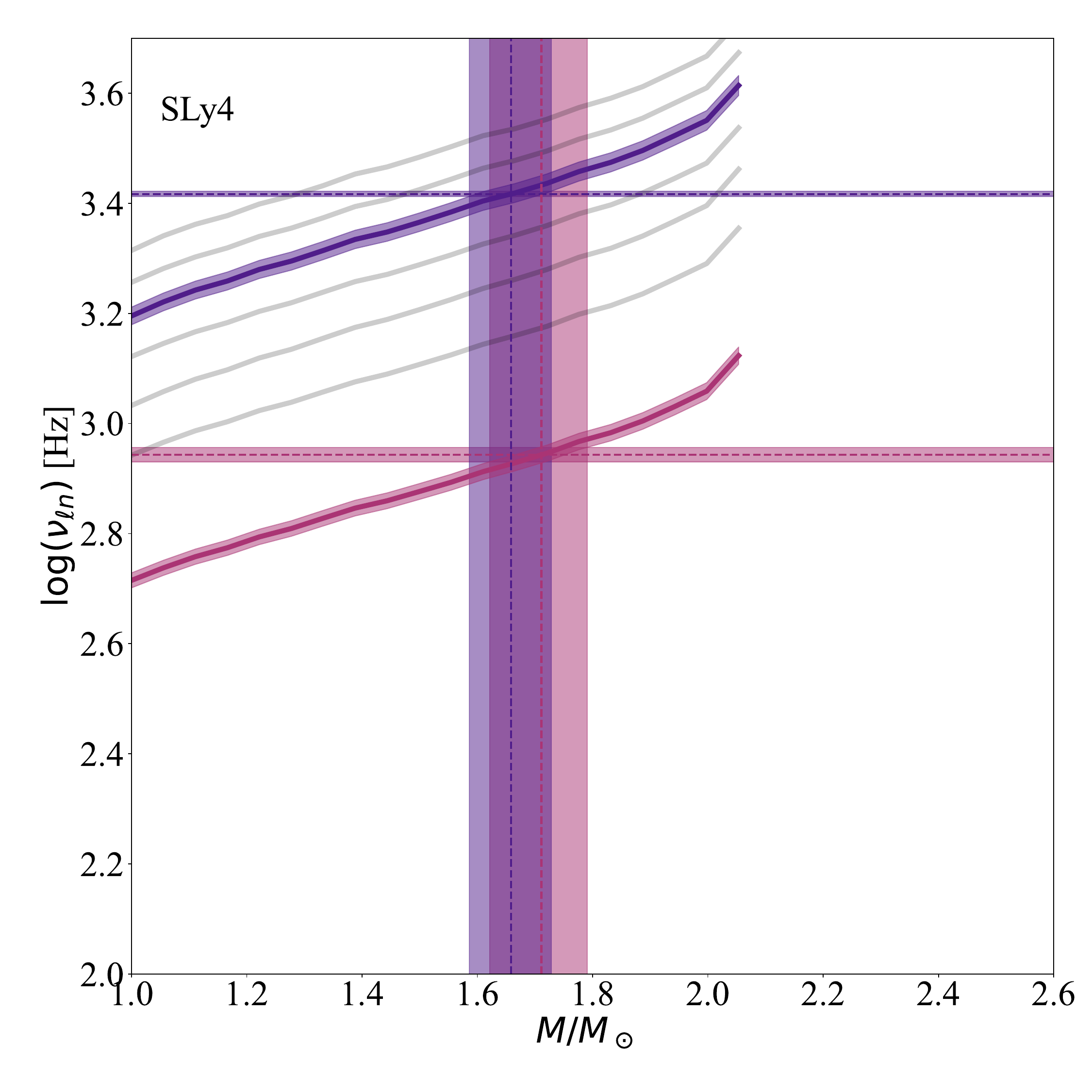}
            \includegraphics[width=0.32\textwidth]{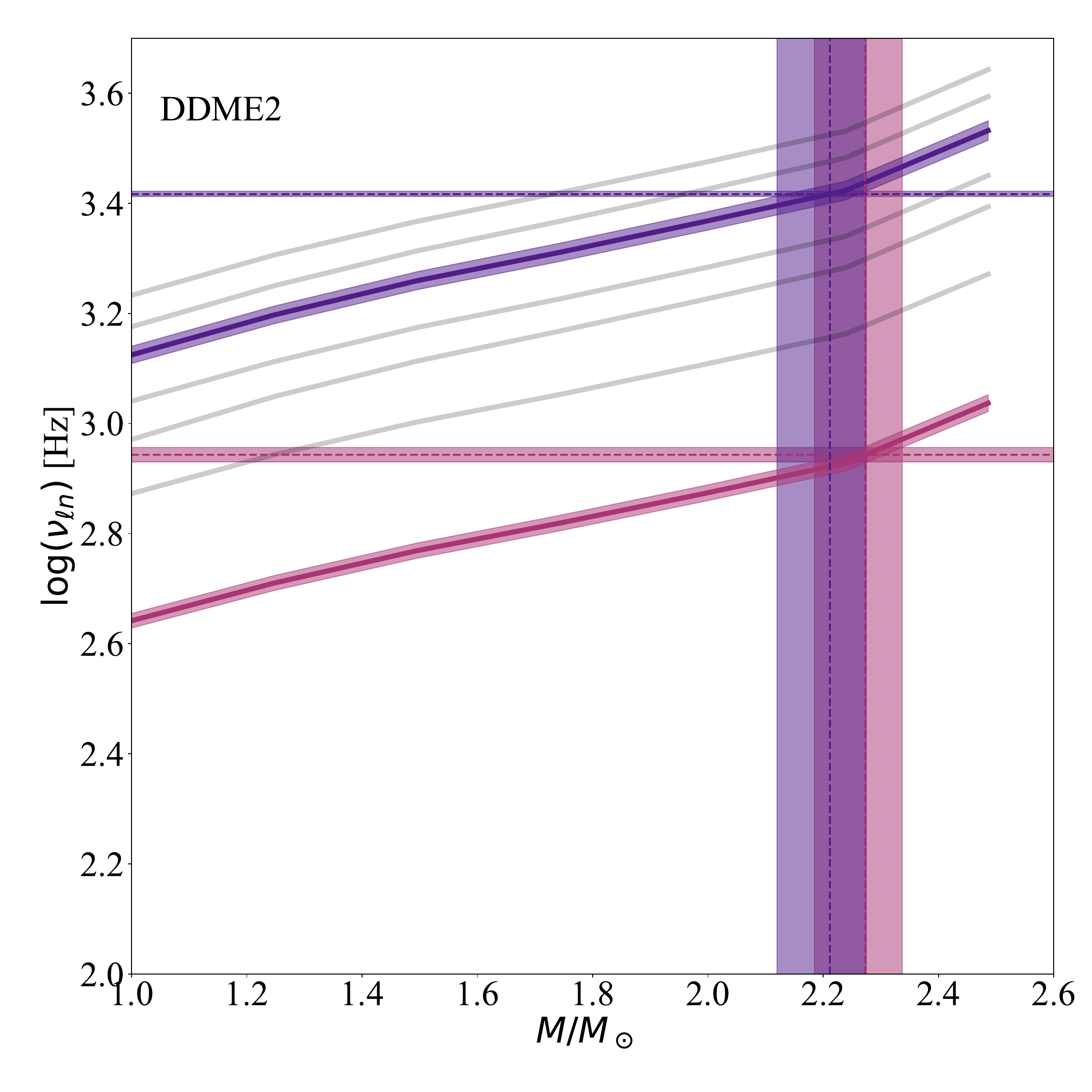}
            \caption{The frequency spectrum $\log\nu_n$ vs. the mass $M$ of an NS. The spectrum lines from the bottom to top correspond to $n=1,2,3,4,5,6,7$. The pink and violet horizontal dashed lines correspond to the two high-frequency QPOs from GRB 931101B (see Table \ref{tab1}), and the vertical dashed lines mark the masses corresponding to these two high-frequency QPOs. The shading indicates the error range.}
            \label{Fig1}
        \end{figure*}

\begin{figure*}
            \centering
            \includegraphics[width=0.32\textwidth]{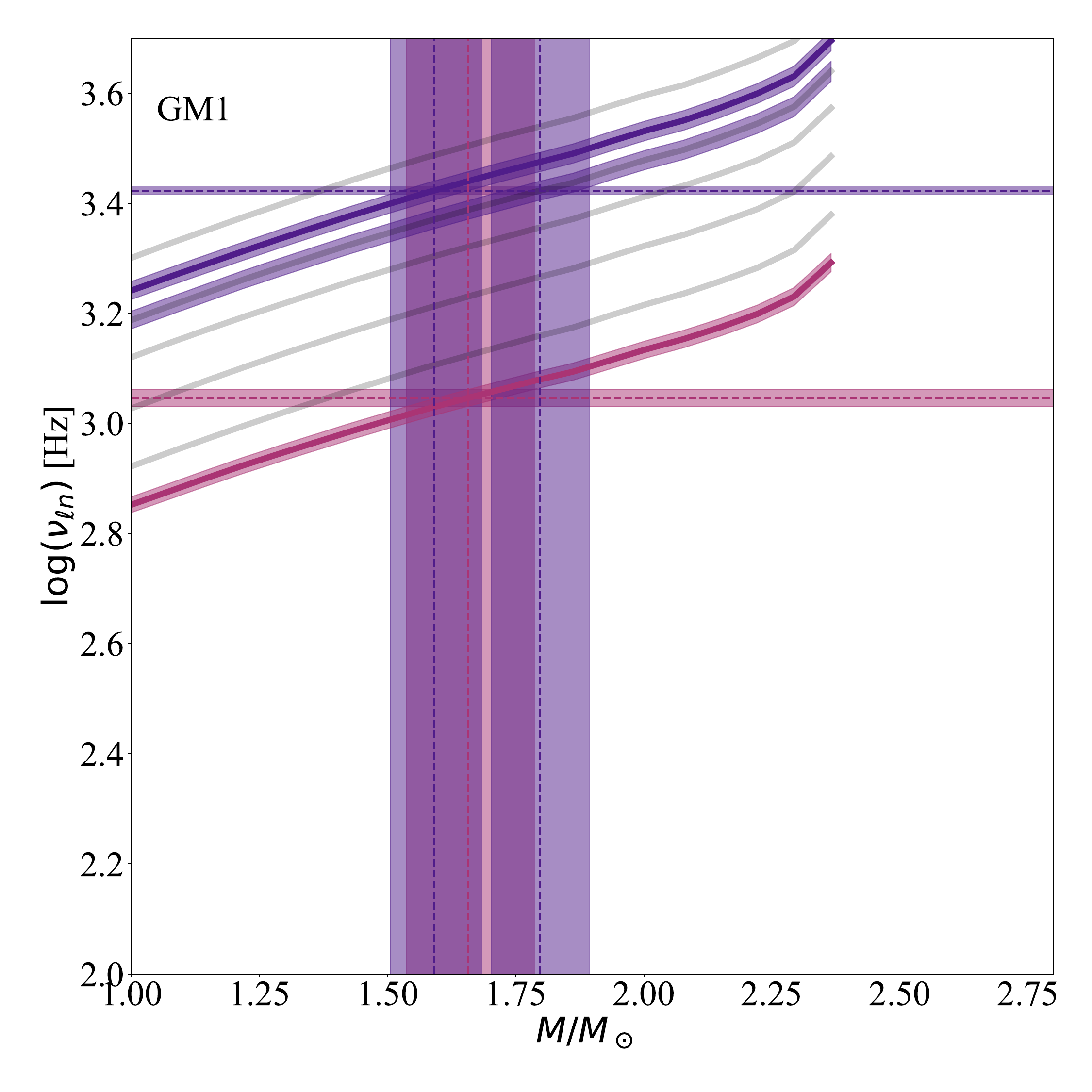}
            \includegraphics[width=0.32\textwidth]{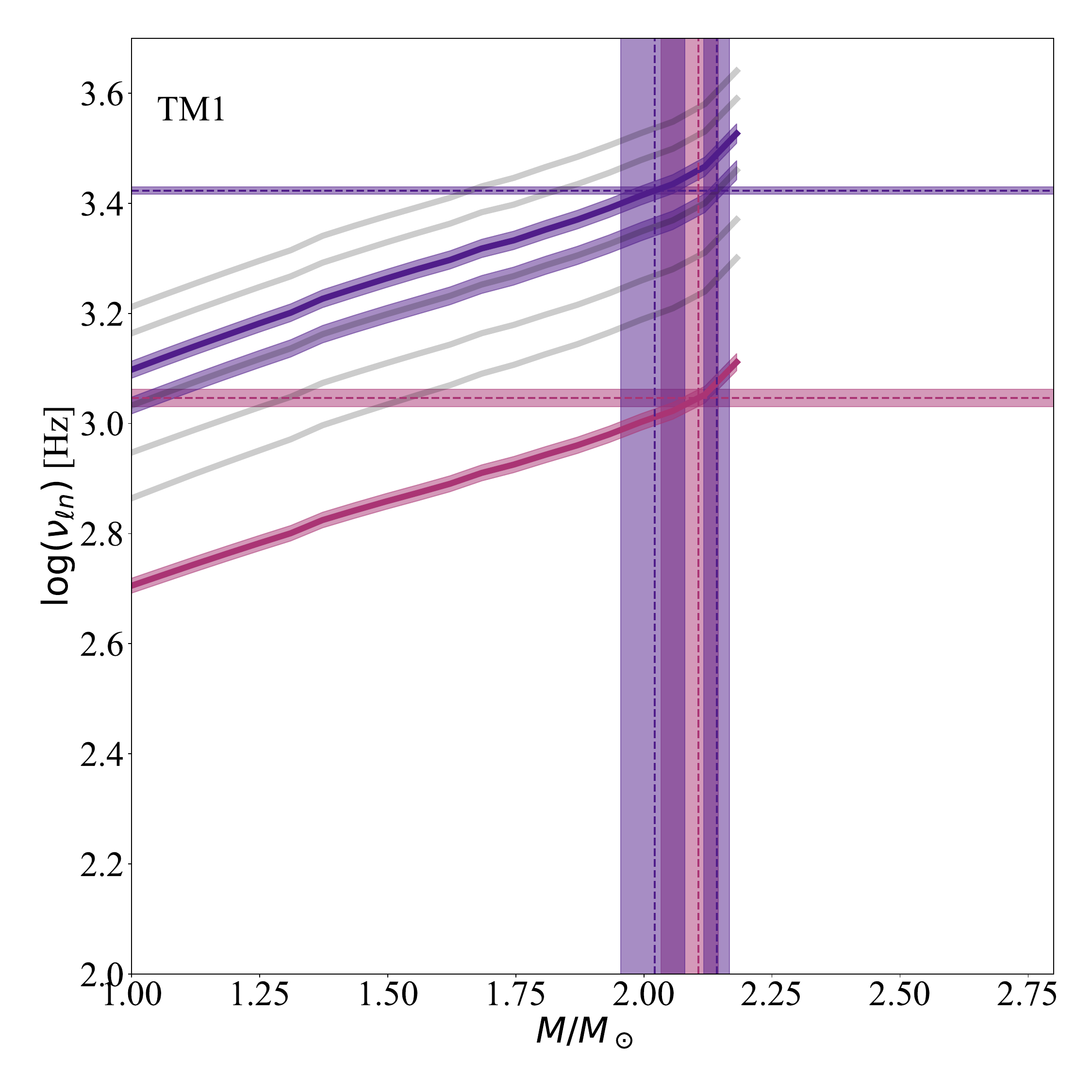}
            \includegraphics[width=0.32\textwidth]{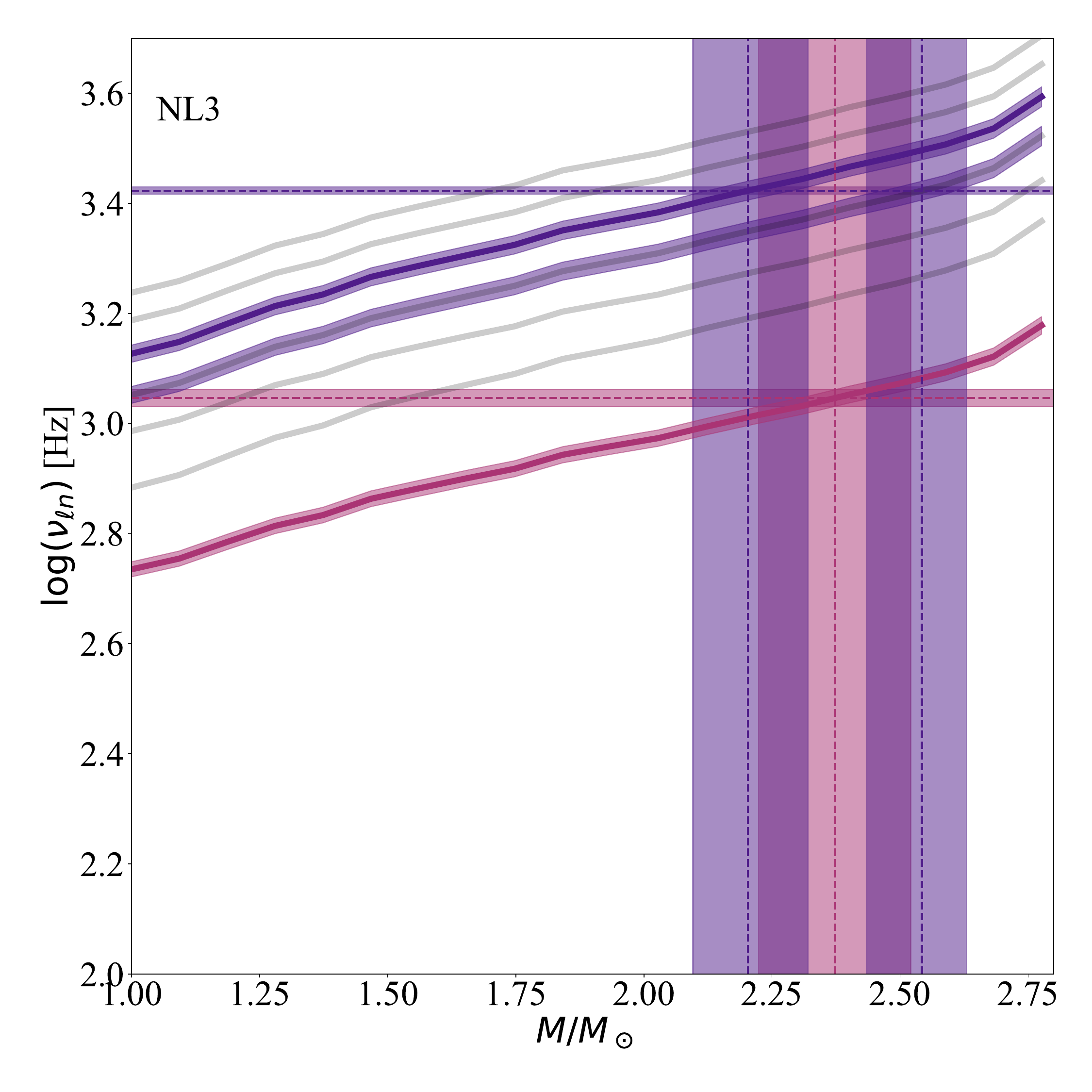}\\
            \includegraphics[width=0.32\textwidth]{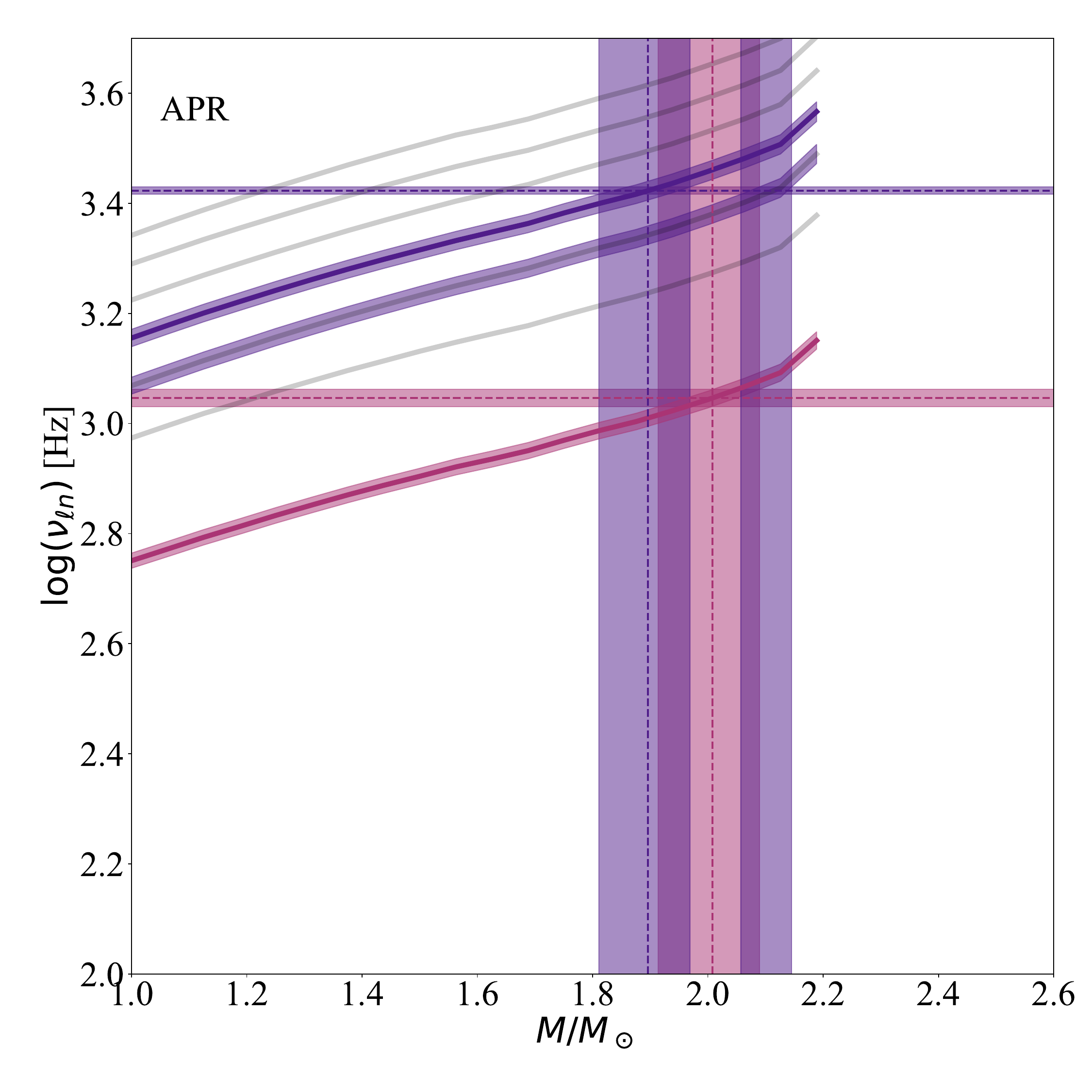}
            \includegraphics[width=0.32\textwidth]{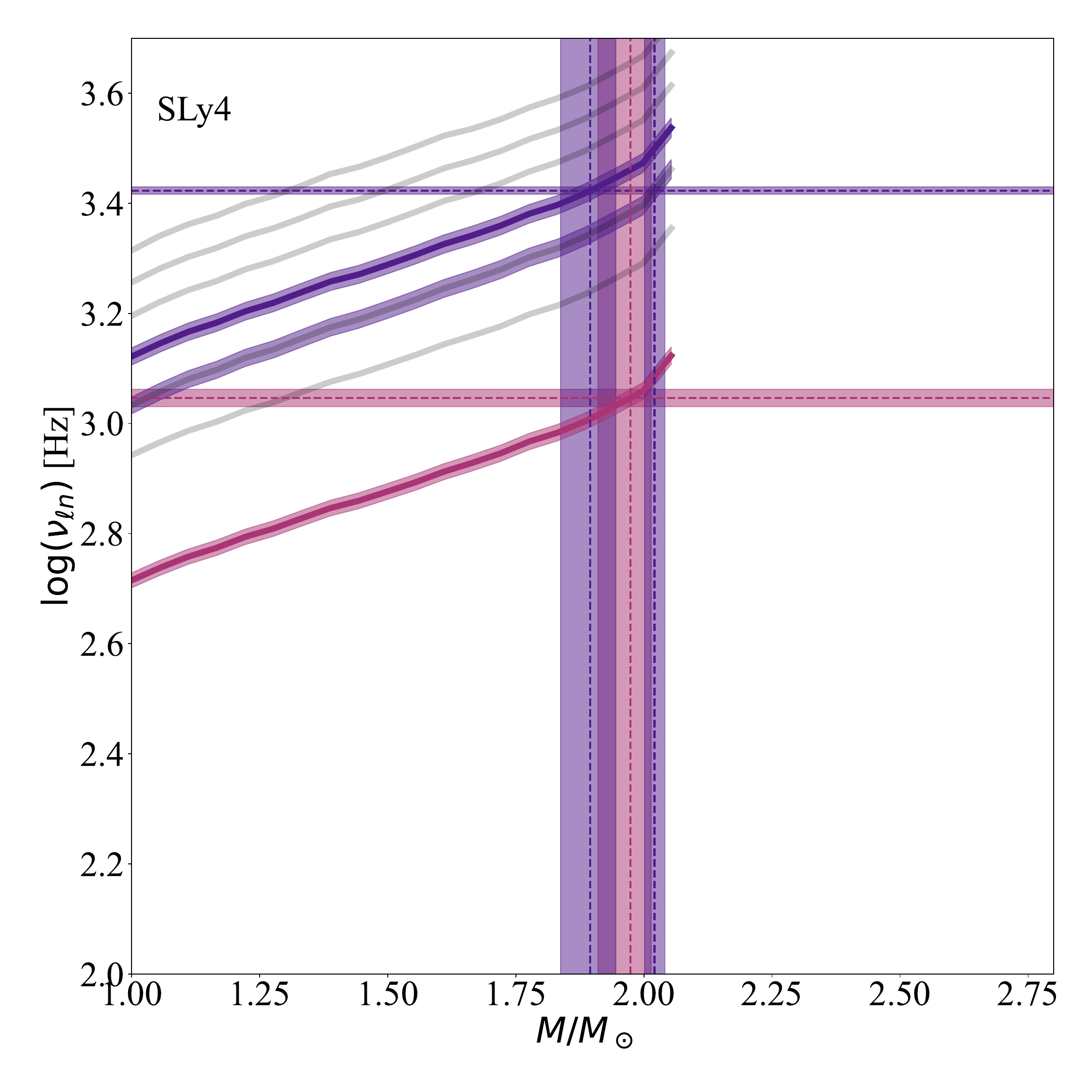}
            \includegraphics[width=0.32\textwidth]{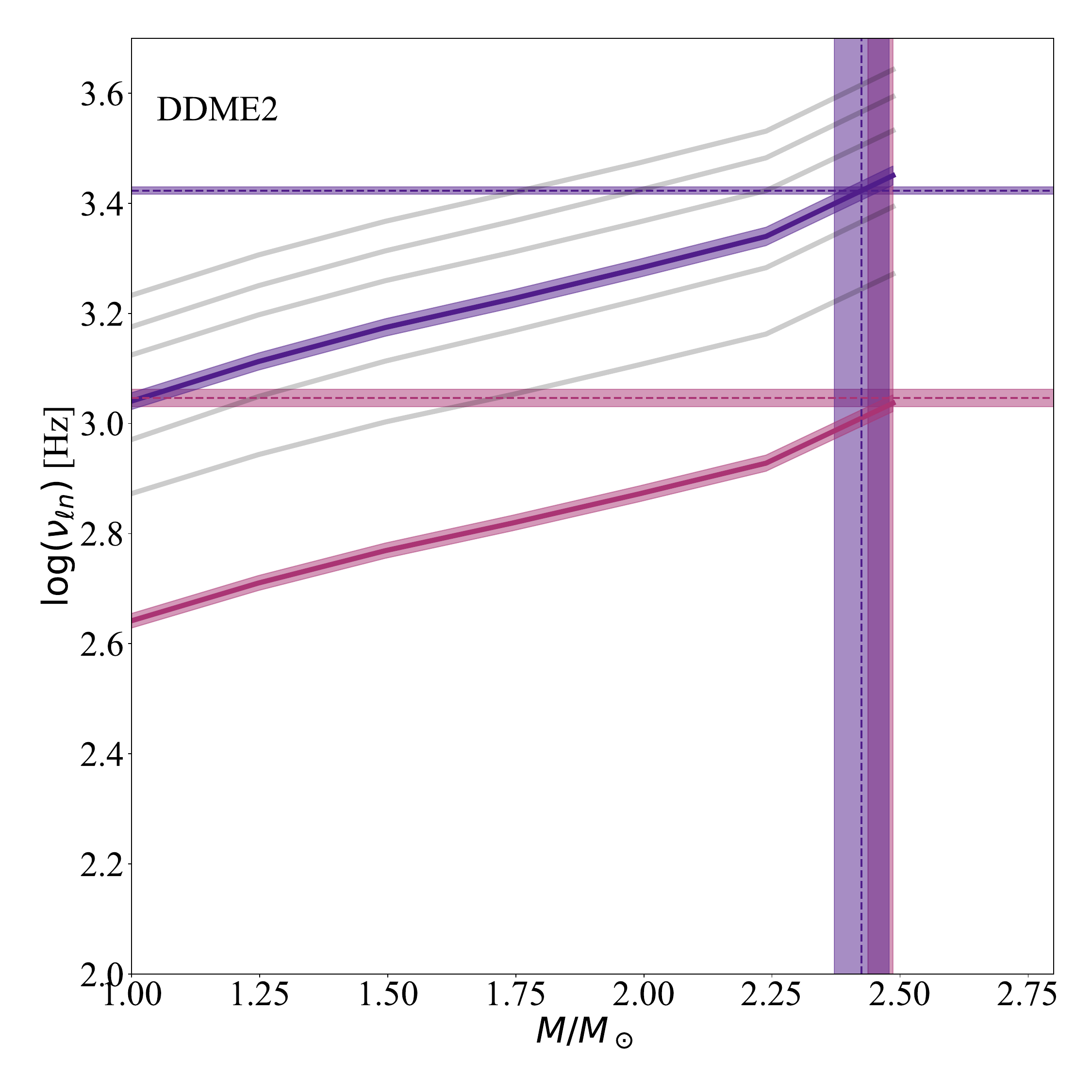}
            \caption{Similar to Figure \ref{Fig1}, but for GRB 910711.}
            \label{Fig2}
        \end{figure*}

        Then, we investigate the relationship between torsional oscillation frequency and mass for different given EOSs of NSs (see Figures~\ref{Fig1} and \ref{Fig2}). In our calculations, we selected six unified EOSs that are suitable for the condition of zero temperature: GM1 \citep{Glendenning.Moszkowski.1991}, TM1 \citep{Sugahara.Toki.1994}, NL3 \citep{Lalazissis.et.al.1997}, APR \citep{Akmal.et.al.1998}, SLy4 \citep{Douchin.Haensel.2001}, and DDME2 \citep{Lalazissis.et.al.2005}. In Figures \ref{Fig1} and \ref{Fig2}, the horizontal dashed lines and shading correspond to the frequencies and error ranges of the two QPOs listed in Table \ref{tab1}, respectively. The vertical dashed lines and shading indicate the ranges of mass that are capable of producing torsional oscillations at those frequencies. Two high-frequency QPOs in the same GRB correspond to two ranges of mass, and it is expected that the ranges of mass for those two high-frequency QPOs can overlap within the margin of error.
        
        By considering the effects of the magnetic field, rotation, and nonspherical symmetry of NS, but with limited knowledge about these factors, we adopt a 0.5\% error in the frequency spectrum lines. For example, the magnetic field confined to the crust may affect the frequencies of various torsional modes. One can make a rough estimate of such an effect of the magnetic field B by using Newtonian physics, which is expressed as \citep{Duncan.1998,Messios.Papadopoulos.2001}
        \begin{equation}
            \frac{_\ell \nu_{n}}{_\ell \nu_{n}^{(0)}}=\left[1+\left(\frac{B}{4 \times 10^{15}}\right)^{2}\right]^{1 / 2},
        \end{equation}
        where $ _\ell \nu_{n}^{(0)}$ is the frequency in the nonmagnetic case. Based on the numerical simulation of torsional and magneto-elastic oscillations of a realistic magnetar model, \citet{Gabler.et.al.2011} found that a $\sim10^{14}$ G magnetic field can achieve a similar damping time on the order of tens of milliseconds. For the given value of magnetic field of $\sim 4 \times 10^{14}$ G, it will result in a correction closer to 0.5\%. This allows us to believe that the 0.5\% error in the frequency spectrum we adopt is reasonable. On the other hand, the nonmeasured redshift ($z$) of those two short-duration GRBs also introduces significant uncertainty. The observed frequency of QPOs is directly related to the frequency in the rest frame, namely,
        \begin{equation}
            \nu_{\rm obs} = \frac{\nu_{\rm rest}}{1 + z}.
        \end{equation}
        The frequency in the rest frame is higher than that in the observed frame with a factor ($1+z$). However, the observed maximum fluxes of GRB 910711 and GRB 931101B are $1.5\times10^{-4}$ $\text{erg cm}^{-2}~\text{s}^{-1}$ and $2.6\times10^{-5}$ $\text{erg cm}^{-2}~\text{s}^{-1}$, respectively. On the contrary, the maximum isotropic energy Eiso in the GF sample can only reach to $5.3\times10^{46}$ erg \citep{Burns.et.al.2021}. By assuming that the isotropic energy can reach to the maximum isotropic energy of GFs from SGRs, one can derive $z < 0.015$, which corresponds to a frequency error of less than 0.15\%. However, it should be emphasized that the error of 0.5\% is only a simple assumption for an uncertainty originating from various complexities. In more extensive error range cases, the ranges of mass required to generate a lower-frequency QPO may overlap simultaneously with multiple ranges of mass (i.e., the intersection of the higher-frequency QPO and spectrum lines with different $n$). This results in two overlapping areas in the mass range, or even the range of masses required for lower-frequency QPOs may be wholly covered. Thus, in this case, the mass range is roughly determined by the QPO with the lower frequency.
        
        For the six given EOSs of NSs in Figures \ref{Fig1} and \ref{Fig2}, it is found that all selected EOSs exhibit significant overlapping mass ranges for both GRB 931101B and GRB 910711. This suggests that an NS with a maximum mass of $\sim\,1.38\ M_{\odot}$ (GM1), $\sim\,1.88\ M_{\odot}$ (TM1), $\sim\,2.01\ M_{\odot}$ (NL3), $\sim\,1.70\ M_{\odot}$ (APR), $\sim\,1.73\ M_{\odot}$ (SLy4), and $\sim\,2.28\ M_{\odot}$ (DDME2) would satisfy the criteria to be considered as a potential source for the QPOs observed in GRB 931101B, and similarly for a maximum mass of $\sim\,1.79\ M_{\odot}$ (GM1), $\sim\,2.15\ M_{\odot}$ (TM1), $\sim\,2.52\ M_{\odot}$ (NL3), $\sim\,2.09\ M_{\odot}$ (APR), $\sim\,2.01\ M_{\odot}$ (SLy4), and $\sim\,2.48\ M_{\odot}$ (DDME2) for GRB 910711 (see Table \ref{tab2}). In other words, we cannot constrain the EOS of NSs via the high-frequency QPOs in these two short-duration GBRs if we believe that they originated from the GFs of magnetars.

    \section{Radial Oscillation}
    The first direct evidence of short GRB origin, which comes from an NS–NS merger, is the detection of GW170817 by Advanced LIGO and Virgo associated with short GRB GRB 170817A and an optical/infrared transient known as kilonova AT 2017gfo \citep{Abbott.et.al.2017, Goldstein.et.al.2017, Zhang.et.al.2018}. From a theoretical point of view, there are four different types of remnants of binary NS mergers, such as a BH, an HMNS, a supramassive NS supported by rigid rotation with a lifetime of hundreds of seconds before collapsing into a BH, and a stable NS \citep{Dai.and.lu.1998a,Dai.and.lu.1998b,Dai.et.al.2006,Rosswog.2000,Fan.Xu.2006,Rezzolla.et.al.2010,Giacomazzo.Perna.2013,Zhang.2013,Lasky.et.al.2014,Lv.et.al.2017,Gao.et.al.2017}. If the remnant is a newborn NS, the temperature is expected to reach several MeV, or even exceed tens of MeV in the immediate aftermath of the merger \citep{Oechslin.et.al.2007,Bauswein.et.al.2010}. Such high temperatures already exceed the point at which a crystalline lattice is stable, and the initial state of the remnant from NS–NS should be fluid filled. The cooling timescale is expected to range from thousands of seconds to days before the crystal forms. Therefore, torsional oscillations, which are shear forces from a crystalline crust, are unlikely to occur, but radial modes are more suitable \citep{Stergioulas.et.al.2011}. The radial oscillation is driven by pressure and gravity, producing a radial displacement that is directly affected by the density and pressure profiles of NSs \citep{S.Chandrasekhar.1964, G.Chanmugam.1977, D.Gondek.1997}. \citet{Gondek.et.al.1997} found that the frequency spectrum of the lowest modes differs significantly from that of cold NSs. In this section, we will discuss the radial oscillation of NSs and constrain the EOS of NSs by adopting the high-frequency QPO of short-duration GRBs GRB 931103B and GRB 910711, assuming that these two short GRBs are indeed from an NS–NS merger.
    
    \subsection{Physical Model}
        Within the framework of general relativity, the study of infinitesimal radial adiabatic oscillations in stars was initially proposed by \cite{S.Chandrasekhar.1964} and applied to enhance its utility for numerical calculation by \cite{G.Chanmugam.1977} later. The study focused on two essential quantities, one being the radial displacement (\(\Delta r\)), which represents the radial motion of a matter element relative to its equilibrium position, and the other being the Lagrangian pressure perturbation (\(\Delta P\)), which quantifies the change in pressure associated with the displacement. These quantities are determined by a pair of ordinary differential equations, which are expressed as follows:
        \begin{align}
        \frac{d\xi}{dr} &= -\frac{1}{r} \left( 3\xi + \frac{\Delta P}{\Gamma P} \right) - \frac{dP}{dr} \frac{\xi}{(P + \rho c^2)}, \label{eq:d_xi_dr} \\
        \frac{d\Delta P}{dr} &= 
        \xi \left\{ \frac{\omega^2}{c^2} e^{\Lambda - \Phi} (P + \rho c^2) r - 4 \frac{dP}{dr} \right\} \notag \\
        &\quad + \xi \left\{ \left( \frac{dP}{dr} \right)^2 \frac{r}{(P + \rho c^2)} - \frac{8\pi G}{c^4} e^{\Lambda} (P + \rho c^2) Pr \right\} \notag \\
        &\quad + \Delta P \left\{ \frac{dP}{dr} \frac{1}{(P + \rho c^2)} - \frac{4\pi G}{c^4} (P + \rho c^2) re^{\Lambda} \right\}. \label{eq:d_deltaP_dr}
        \end{align}
        Here \(\xi = \Delta r /r\) is the normalized radial displacement. To avoid the singularity in \(\xi\) at the stellar center and ensure a wellbehaved normalization for the eigenfunctions, we adopt \(\xi(0) = 1\) as the boundary condition. Parameter \(\omega_n =2\pi \nu_n\) is the angular frequency of radial oscillation with the eigenfrequency $\nu_n$; the stability condition of oscillations requires \(\omega^2 \ge 0\). $\Gamma$ is the adiabatic index that is related to the baryon number density \(n_b\) and pressure \citep{BPS.1971}, namely,
        \begin{equation}
        \Gamma=\frac{n_b}{P}\frac{dP}{dn_b}=\frac{P + \rho c^2}{P c^2} \frac{dP}{d\rho} 
        \end{equation}
        
        By solving Equations (\ref{eq:d_xi_dr}) and (\ref{eq:d_deltaP_dr}), it is necessary to apply boundary conditions at the stellar center and surface, namely,
        \begin{equation}
        \Delta P = -3 \xi \Gamma P, \quad \text{for}~r = 0. 
        \end{equation}
        \begin{equation}
        \Delta P = 0, \quad \text{for}~r = R.
        \end{equation}
        If this is the case, Equations (\ref{eq:d_xi_dr}) and (\ref{eq:d_deltaP_dr}) can be reduced to a second-order differential equation for \(\xi(r)\) and switched to the form of a Sturm–Liouville eigenvalue problem. Similar to the case of torsional oscillation, the eigenvalues \(\omega^2\) can also yield a discrete spectrum \(\omega_0^2 < \omega_1^2< \dots<\omega_n^2\), and each eigenvalue corresponds to an eigenfunction \(\xi_n\) that has \(n\) nodes within \(0 \leq r \leq R\). The fundamental mode (\(n = 0\)), also known as the $f$-mode, corresponds to the lowest eigenvalue, while the higher modes with \(n = 1, 2, 3, \dots\) are referred to as $p$-modes \citep{Sagun.2020}. The eigenfrequencies not only describe the nature of stellar oscillations but also serve as critical indicators of the system’s stability under small perturbations. Hence, all eigenvalues should satisfy \(\omega_n^2 \ge 0\) for every \(n\) and ensure that a static stellar model is dynamically stable under small, radial, and adiabatic perturbations.

        \subsection{Constraint on the EOS of NS}
        An HMNS, supported by differential rotation that survives 10–100 ms before collapsing into a BH, is a possible remnant of a binary NS merger. Initially, the HMNS is expected to have an extremely high temperature, potentially reaching tens of MeV. Such high temperatures significantly affect the physical properties of HMNSs, such as the EOS of NSs. By assuming that the high-frequency QPOs in the short-duration GRBs GRB 931101B and GRB 910711 are caused by the radial oscillation of an HMNS from a binary NS merger, we employ six EOSs for NSs (NL3, IUF, TM1, TMA, FSG, and BHBLp), which are suitable for the condition of high temperature. The interpolated data for the selected EOS models were obtained from the publicly available CompOSE database\footnote{\url{https://compose.obspm.fr/table}.} \citep{Oertel:2016bki,Typel:2013rza}.

\begin{figure*}
            \centering
            \includegraphics[width=0.32\textwidth]{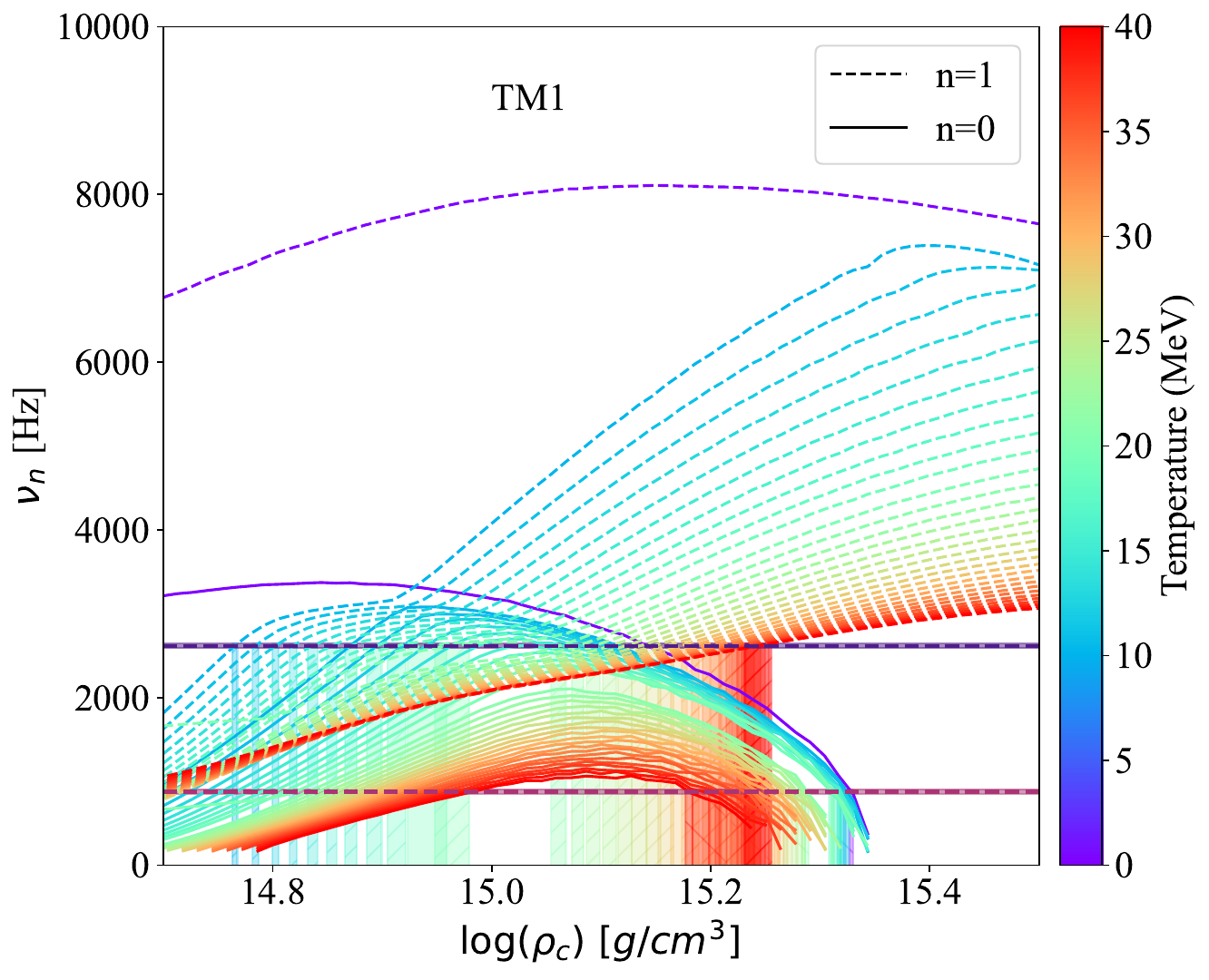}
            \includegraphics[width=0.32\textwidth]{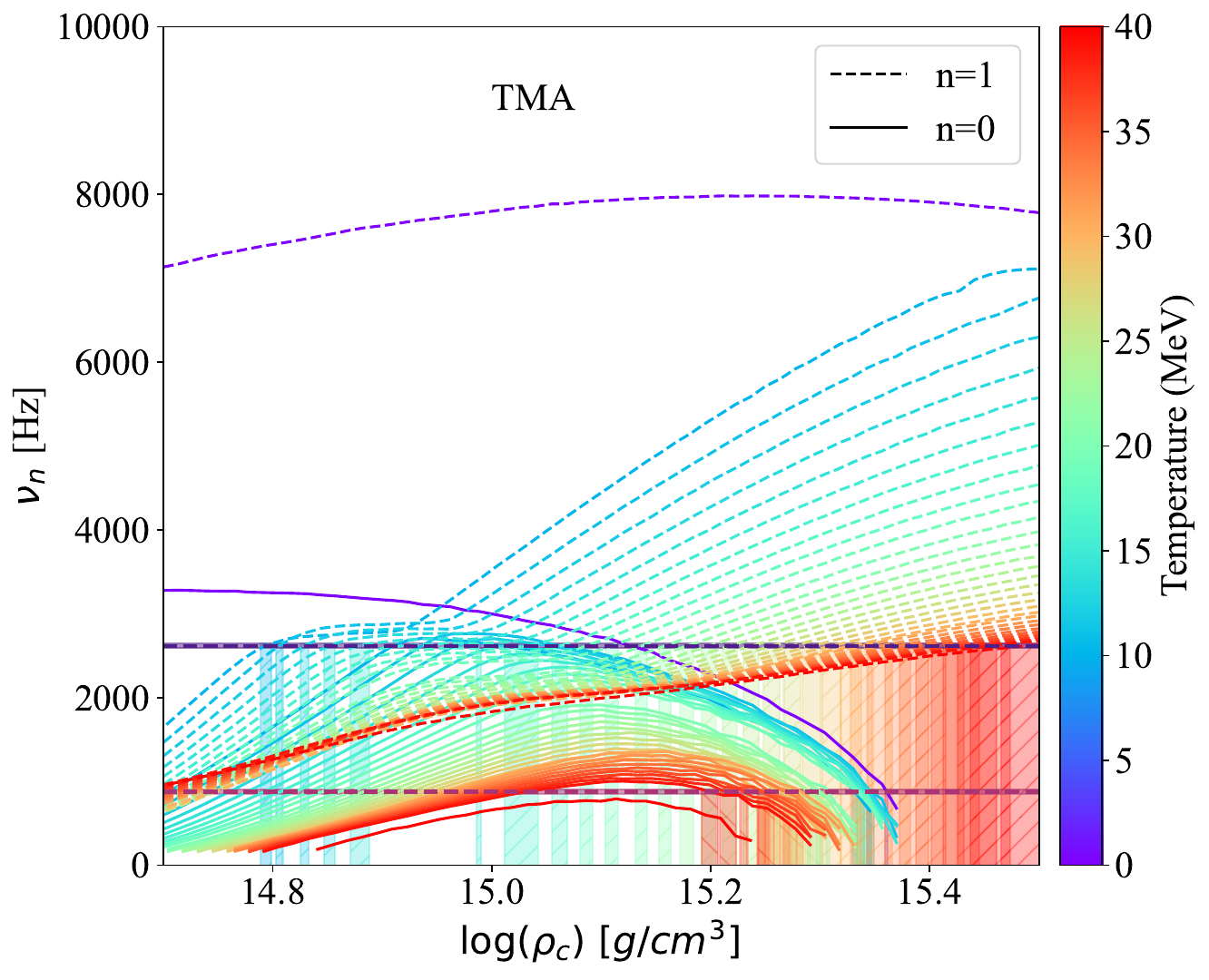}
            \includegraphics[width=0.32\textwidth]{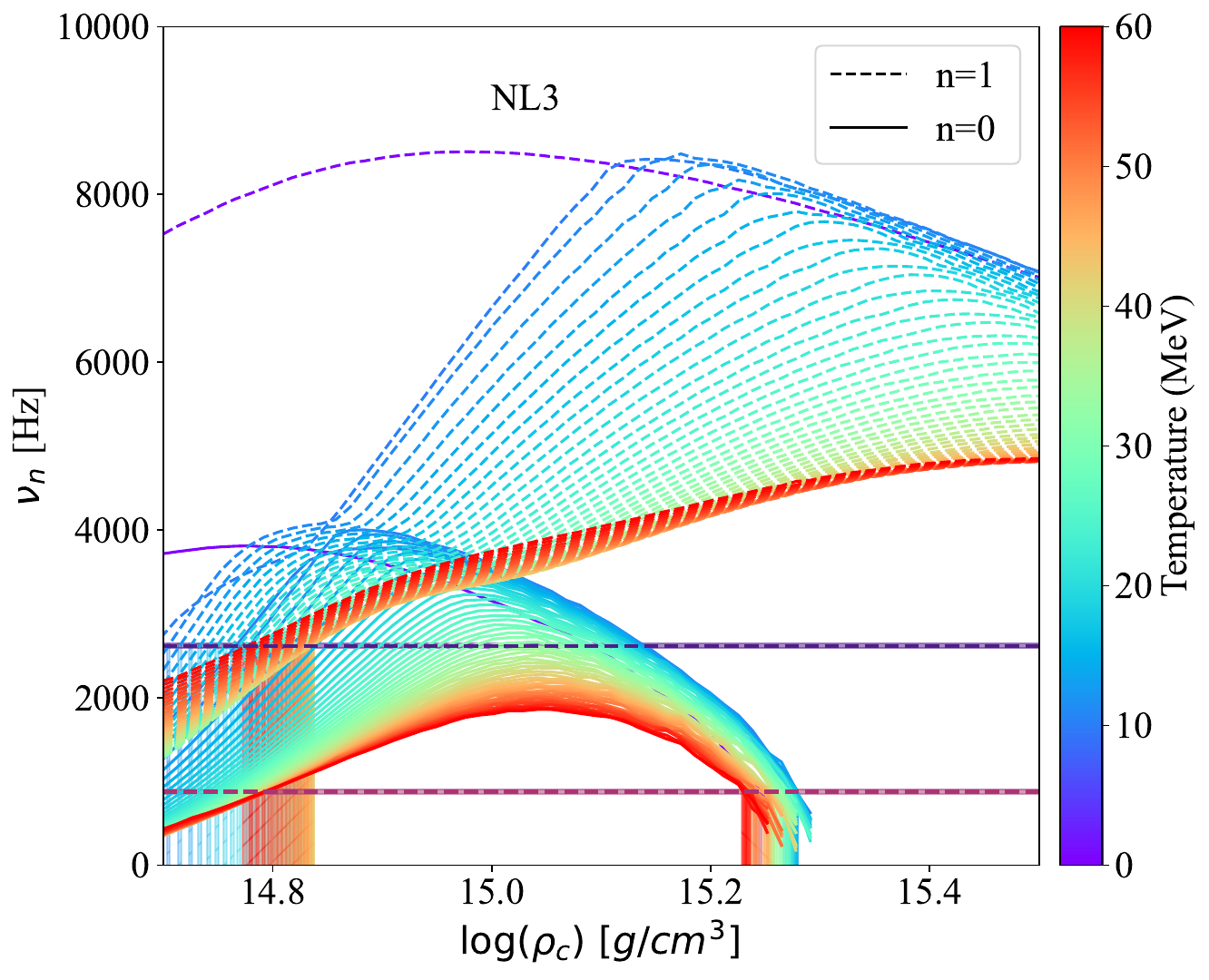}\\
            \includegraphics[width=0.32\textwidth]{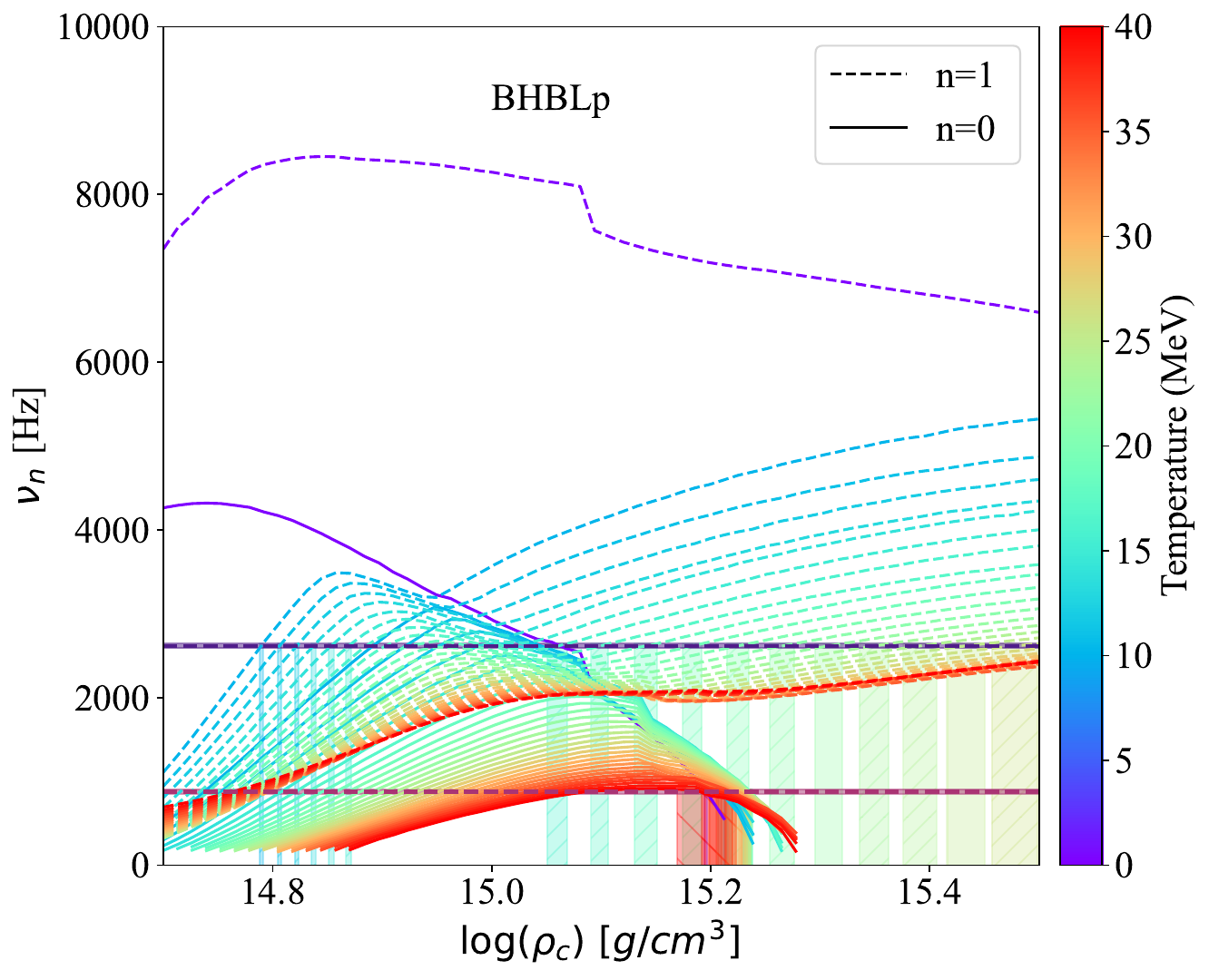}
            \includegraphics[width=0.32\textwidth]{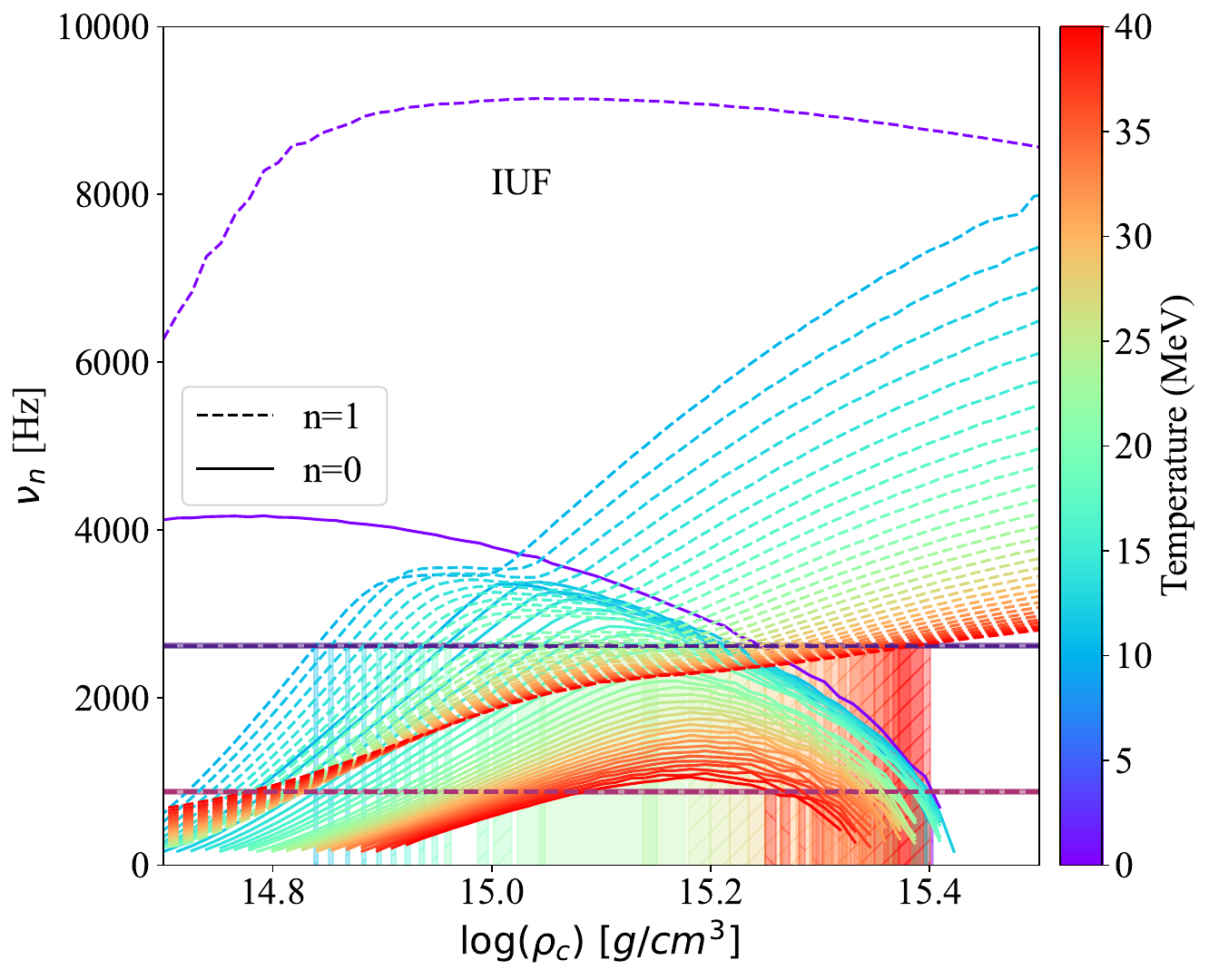}
            \includegraphics[width=0.32\textwidth]{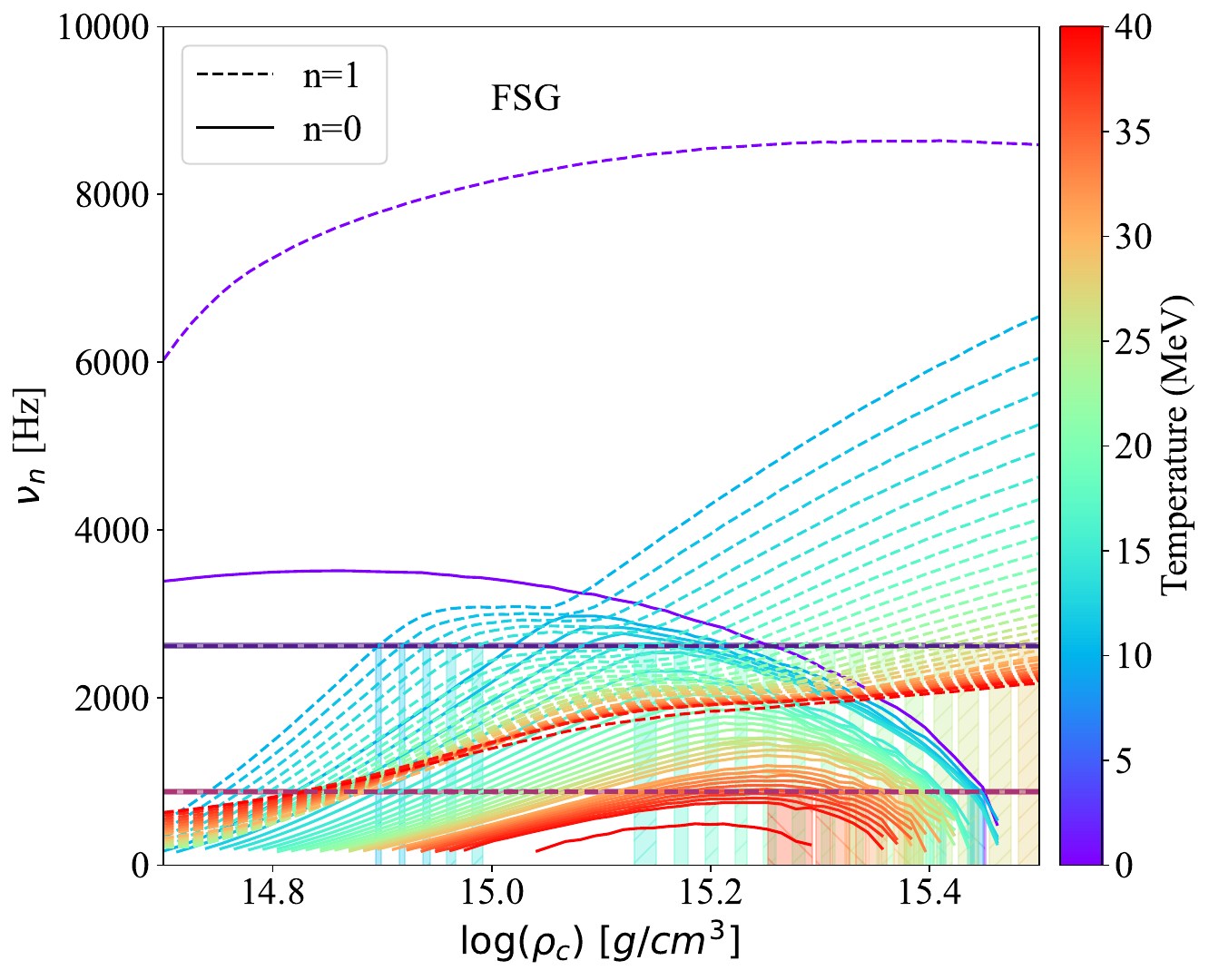}
            \caption{Eigenfrequencies ($\nu_n$) of various EOS models as a function of the central density ($\log(\rho_c)$) for different temperatures. The solid and dashed lines correspond to $n=0$ modes (i.e., $f$-mode) and $n=1$ modes (i.e., $p$-mode), respectively. The color bar is the temperature in units of MeV. The horizontal lines represent the two observed high-frequency QPOs for GRB 931101B. The shaded vertical regions correspond to the ranges of $\rho_c$ where the eigenfrequencies of $n=0$ and $n=1$ modes match with the observed lower and higher frequencies of the QPO, respectively.}
            \label{Fig3}
        \end{figure*}

\begin{figure*}
            \centering
            \includegraphics[width=0.32\textwidth]{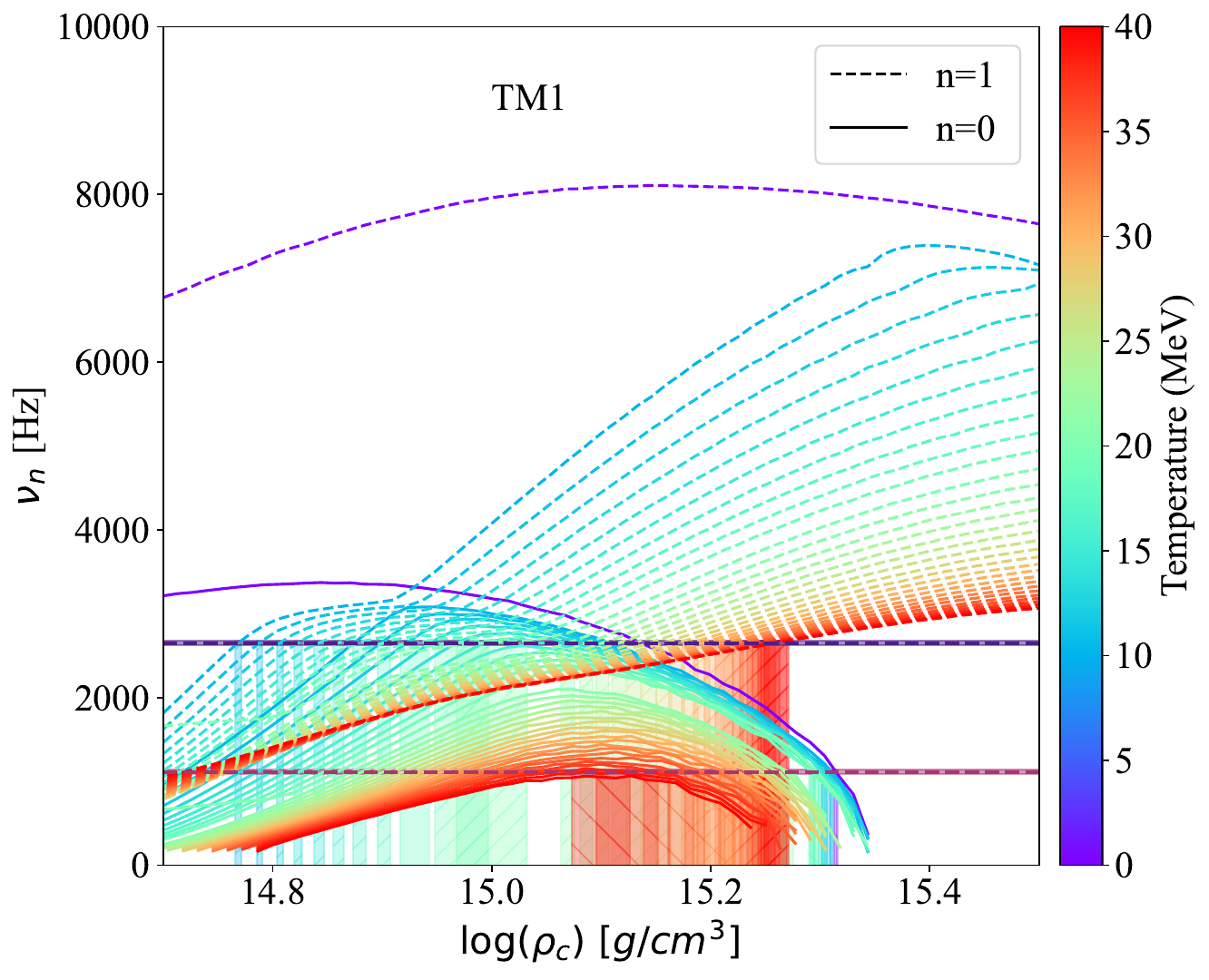}
            \includegraphics[width=0.32\textwidth]{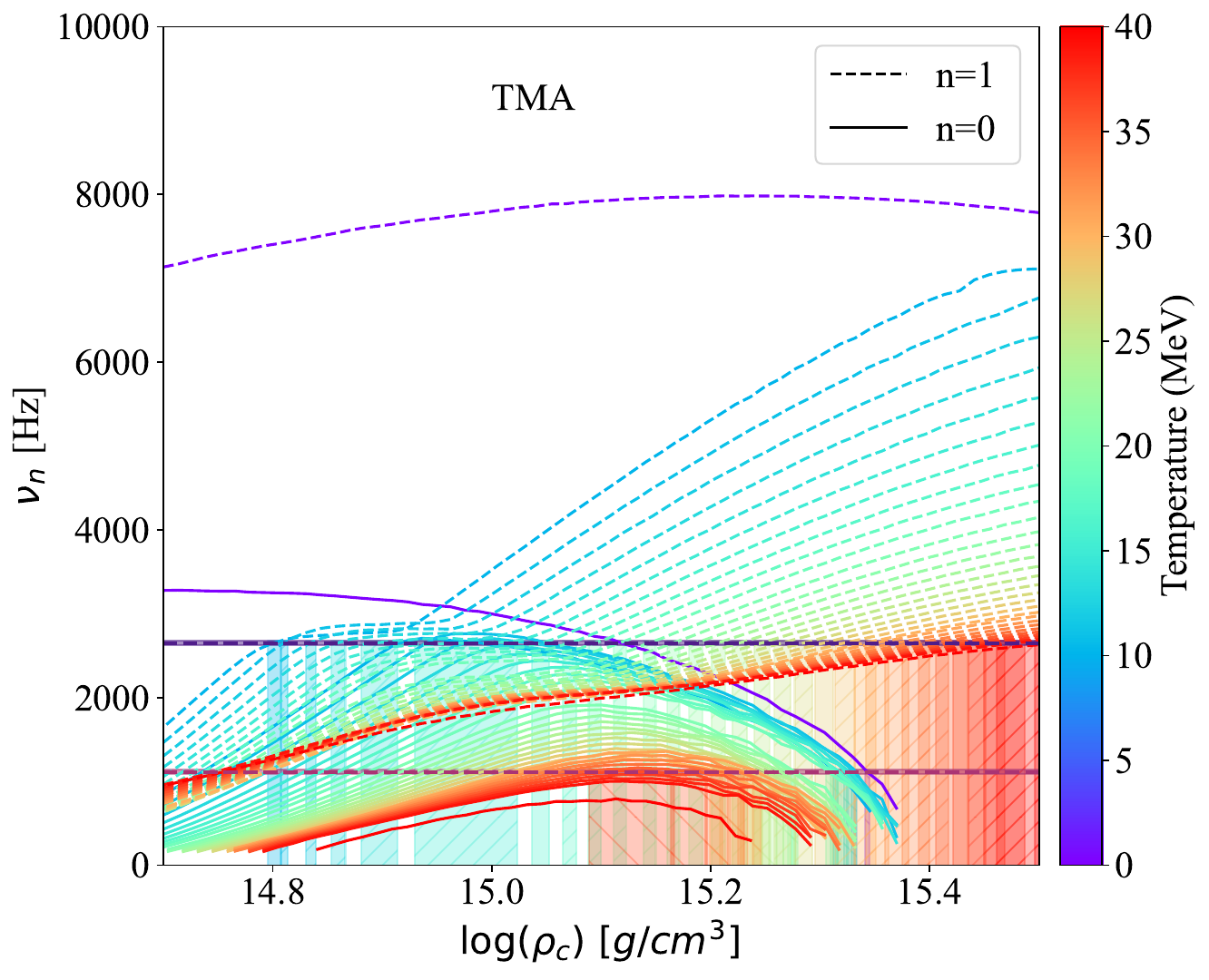}
            \includegraphics[width=0.32\textwidth]{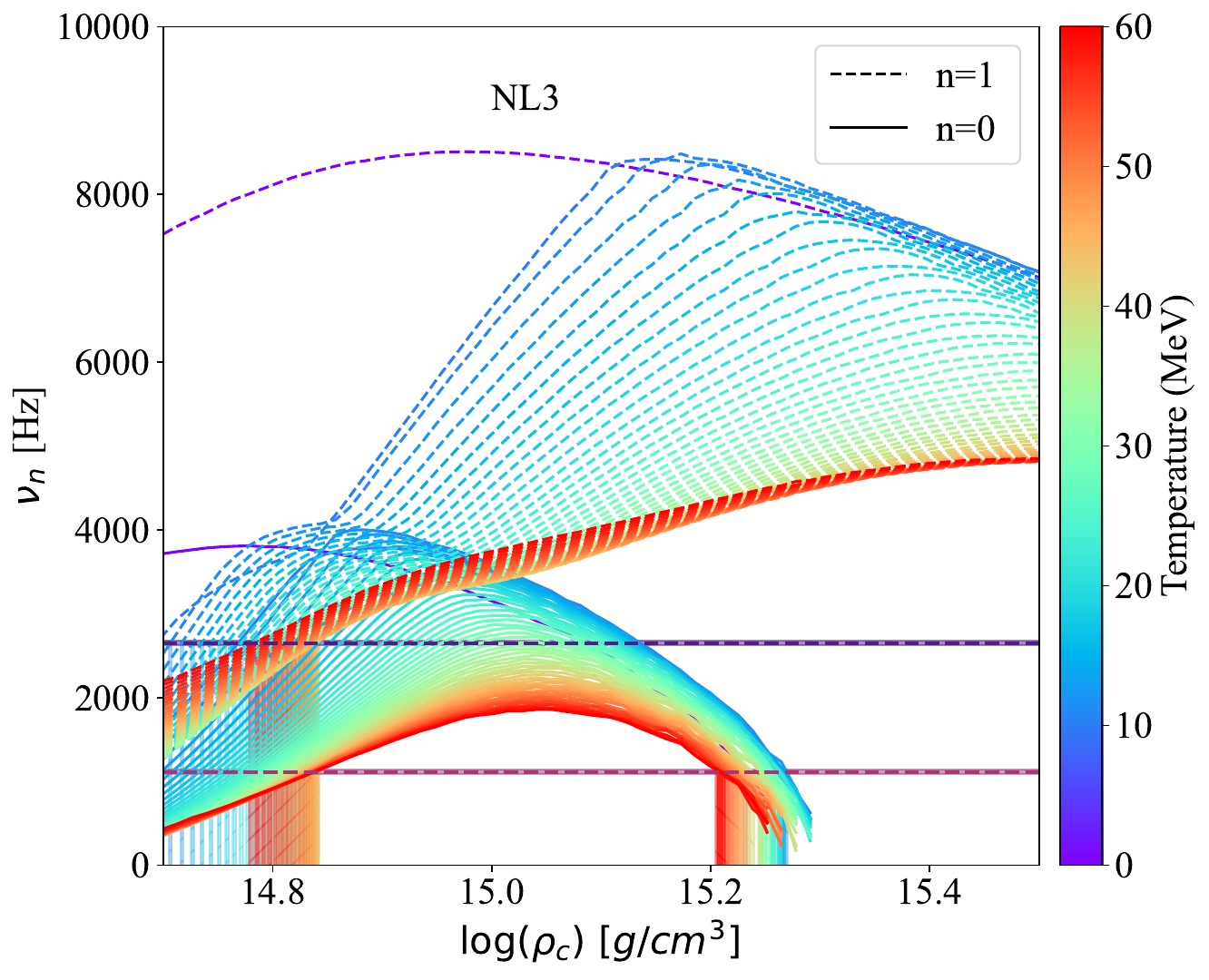}\\
            \includegraphics[width=0.32\textwidth]{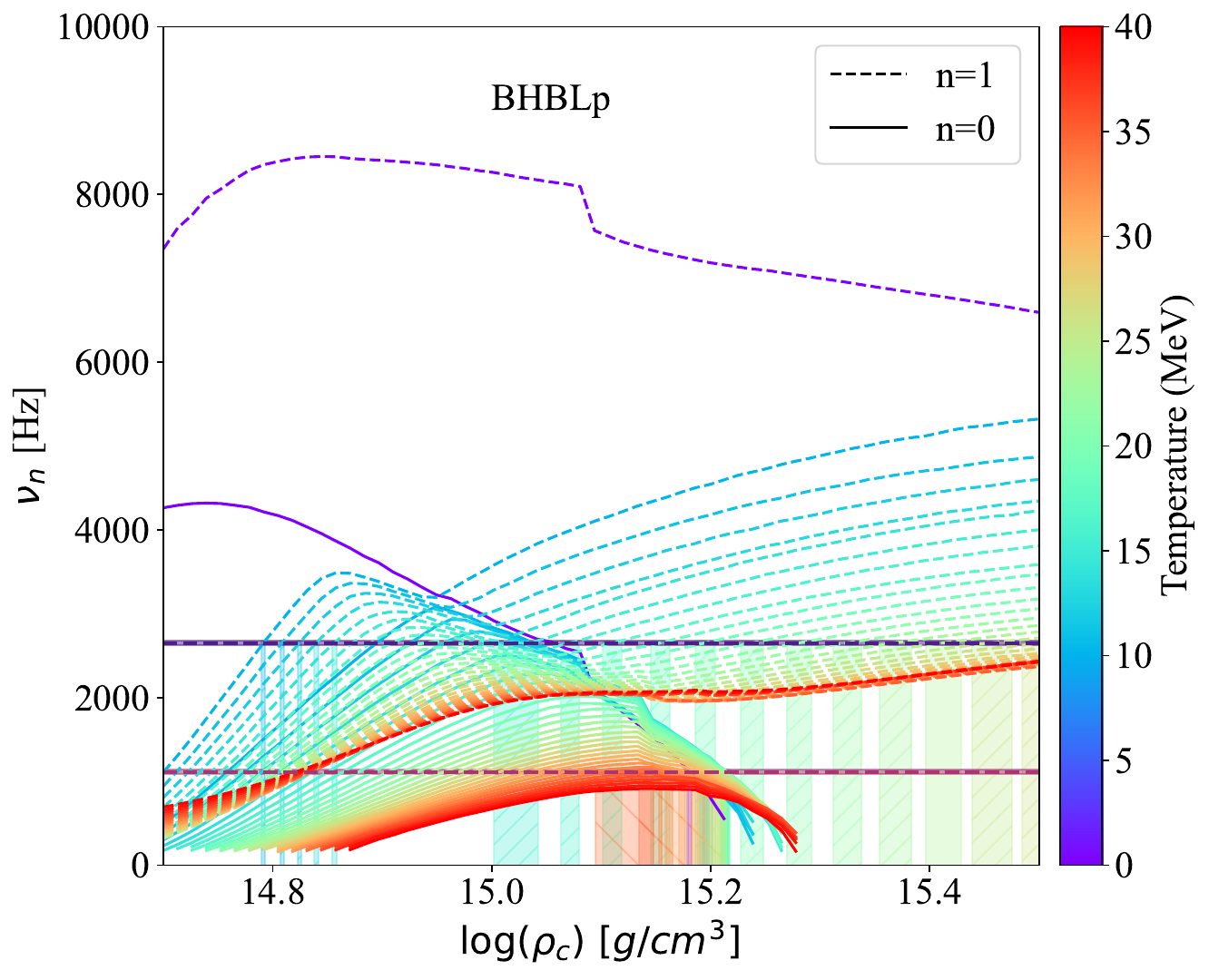}
            \includegraphics[width=0.32\textwidth]{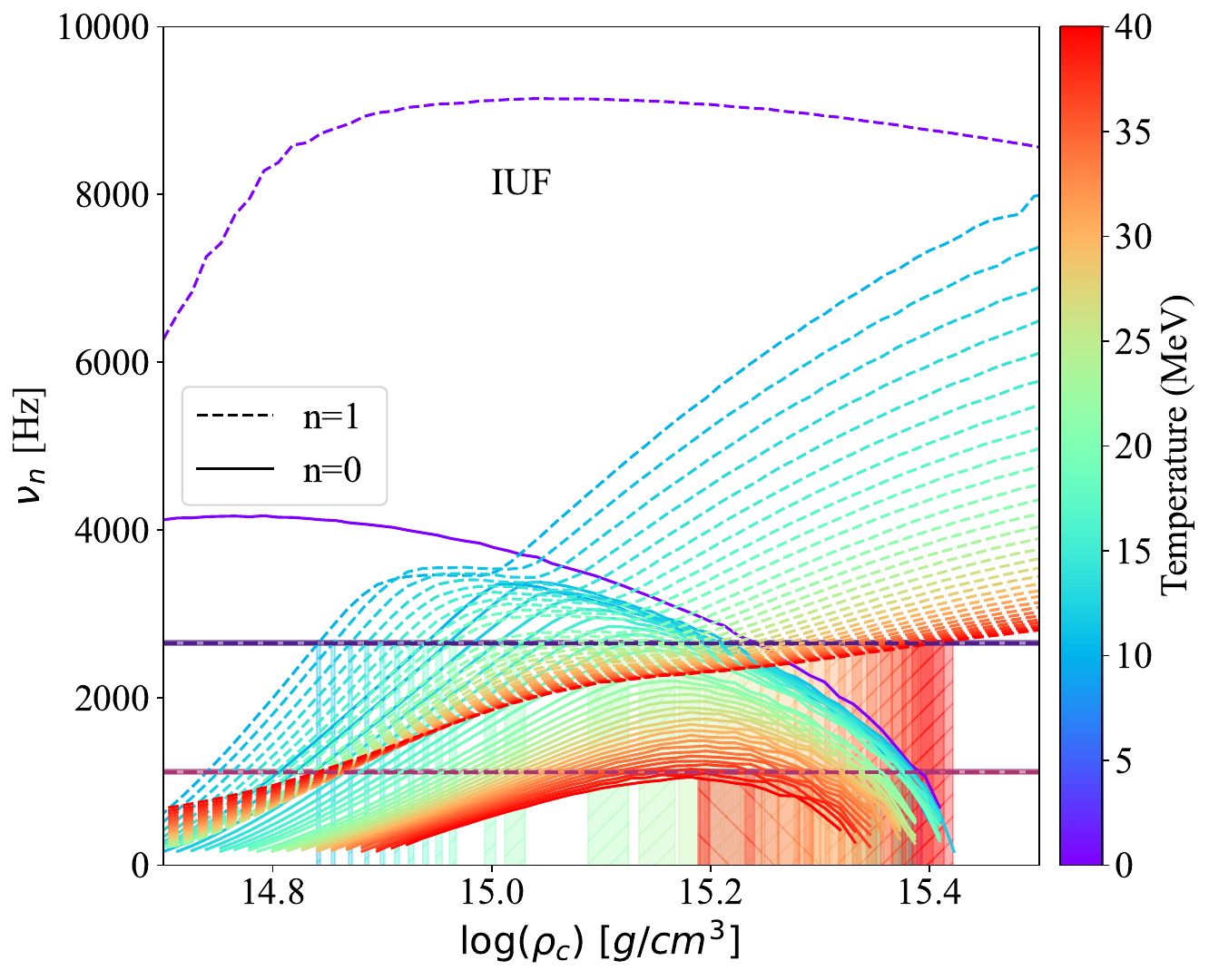}
            \includegraphics[width=0.32\textwidth]{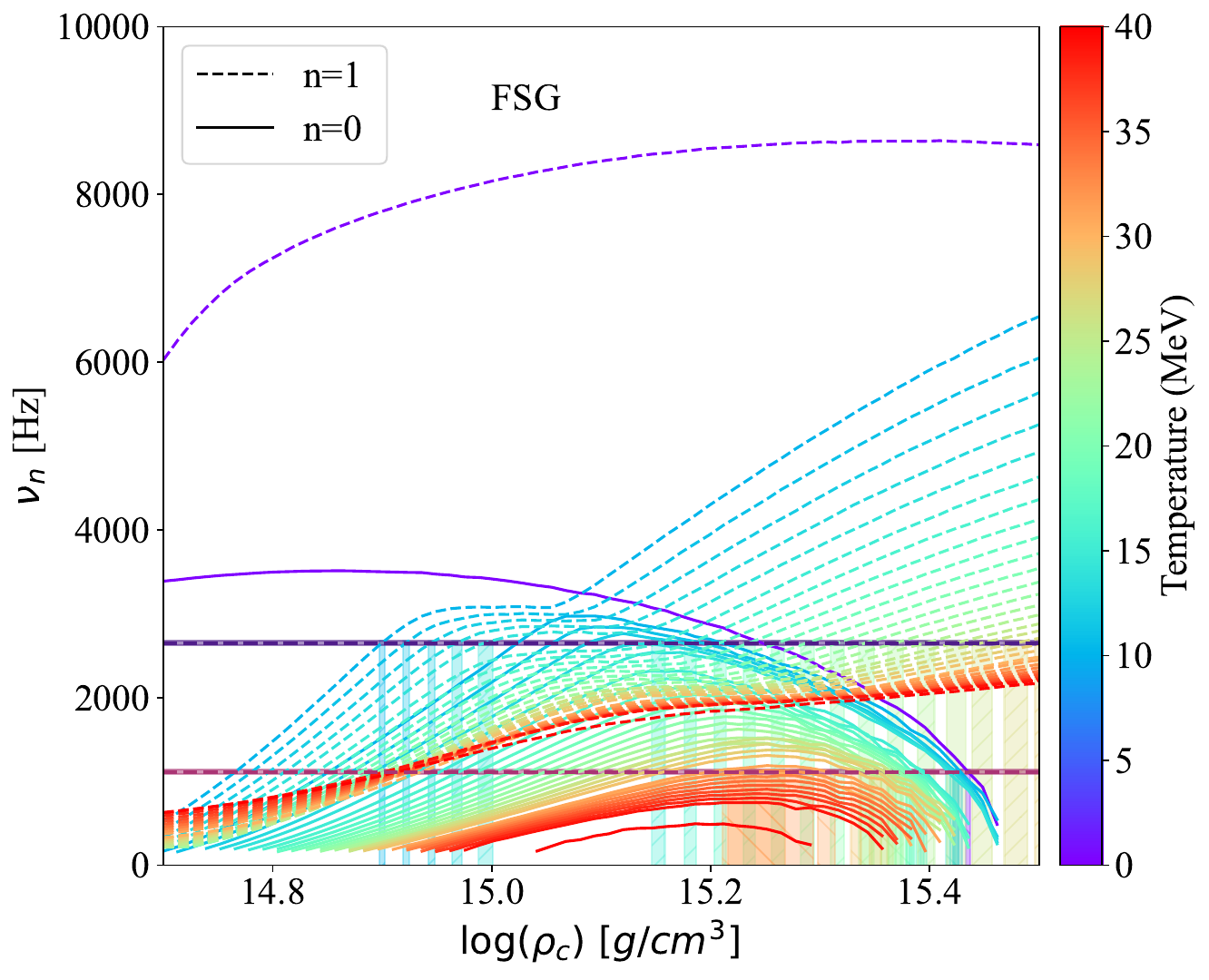}
            \caption{Similar to Figure \ref{Fig3}, but for GRB 910711.}
            \label{Fig4}
        \end{figure*}

        Figures \ref{Fig3} and \ref{Fig4} show the eigenfrequencies ($\nu_n$) of various EOS models as functions of the central density ($\rm log \rho_c$) with different temperatures for GRB 931101B and GRB 910711, respectively. The maximum temperature of five EOSs (i.e., IUF, TM1, TMA, FSG, and BHBLp) can reach 40 MeV, while for NL3 it extends up to as high as 60 MeV. It is found that the frequencies of \(n=1\) modes at zero temperature are significantly higher than those of observed values for those two short GRBs. However, the frequency of the \(n=1\) mode is rapidly decreased when the temperature is increased. The shaded vertical regions in Figures \ref{Fig3} and \ref{Fig4} correspond to the ranges of $\rho_c$ where the eigenfrequencies of \(n=0\) and \(n=1\) modes match the observed lower and higher frequencies of the QPO, respectively. If this is the case, the temperature at which the observed lower and higher frequencies of the QPO correspond to the same $\rho_c$ needs to be determined and can be used to constrain the EOS of NSs.

\begin{figure*}
            \centering
            \includegraphics[width=0.49\textwidth]{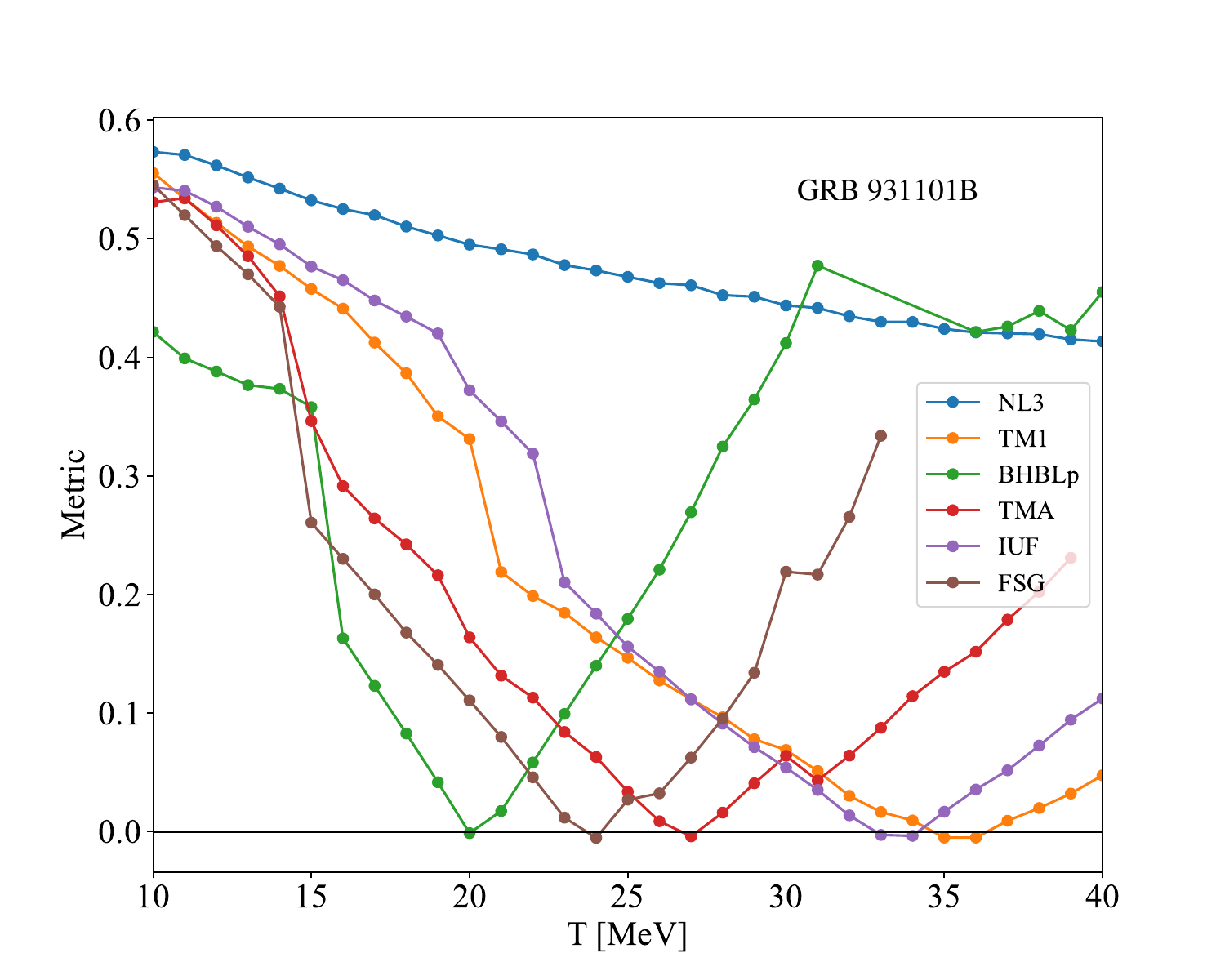}
            \includegraphics[width=0.49\textwidth]{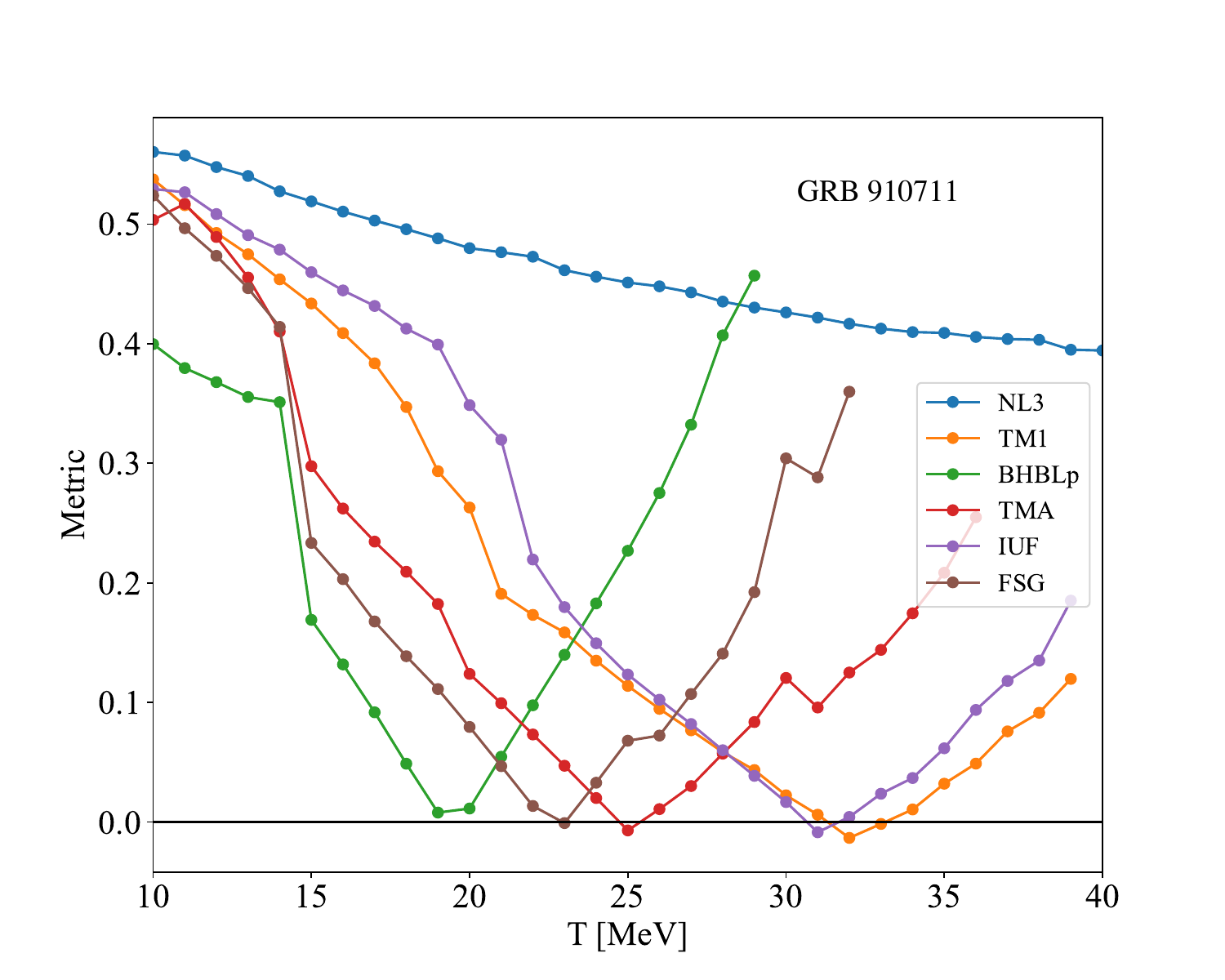}
            \caption{
            Metric obtained for different EOS models with varying temperatures. The Metric quantifies the spatial relationship between the $\rho_c$ ranges corresponding to the $n=0$ and $n=1$ modes.}
            \label{Fig5}
        \end{figure*}

        To find the temperature at which the vertical bands overlap, we define a metric: $\text{Metric}=\rho_{\rm gap}-\rho_{\rm cross}$, which quantifies the spatial relationship between the two vertical bands. Here the $\rho_{\rm gap}$ represents the distance between the closest edges of the two regions, namely,
        \[
        \rho_{\rm gap}= \max[0, \max(x_{0,\text{min}}, x_{1,\text{min}}) - \min(x_{0,\text{max}}, x_{1,\text{max}})]
        \]
        Here \([x_{0,\text{min}}, x_{0,\text{max}}]\) and \([x_{1,\text{min}}, x_{1,\text{max}}]\) are the $\rho_c$ ranges obtained by comparing the $n=0$ mode with the observed lower-frequency QPO and the $n=1$ mode with the higherfrequency QPO, respectively. If the bands overlap, $\rho_{\rm gap}=0$. Parameter $\rho_{\rm cross}$ is defined as
        \[
        \rho_{\rm cross}= \max[0, \min(x_{0,\text{max}}, x_{1,\text{max}}) - \max(x_{0,\text{min}}, x_{1,\text{min}})]
        \]
        If the bands do not overlap, $\rho_{\rm cross}<0$; otherwise, it is the size of the overlapping region between the two bands. $\text{Metric}=0$ implies that there exists a temperature at which the observed lower and higher frequencies of the QPO correspond to the same $\rho_c$ at a specific temperature. Figure \ref{Fig5} shows the relationship between the Metric and temperature ($T$) for different EOSs. It is found that most EOSs allow the Metric to reach a negative value at a specific temperature for those two short GRBs, except for the stiffest EOS, NL3. Hence, except NL3, we cannot rule out the other five EOSs based on the observed lower and higher frequencies of the QPO for those two short GRBs.

\begin{figure*}
            \centering
            \includegraphics[width=0.49\textwidth]{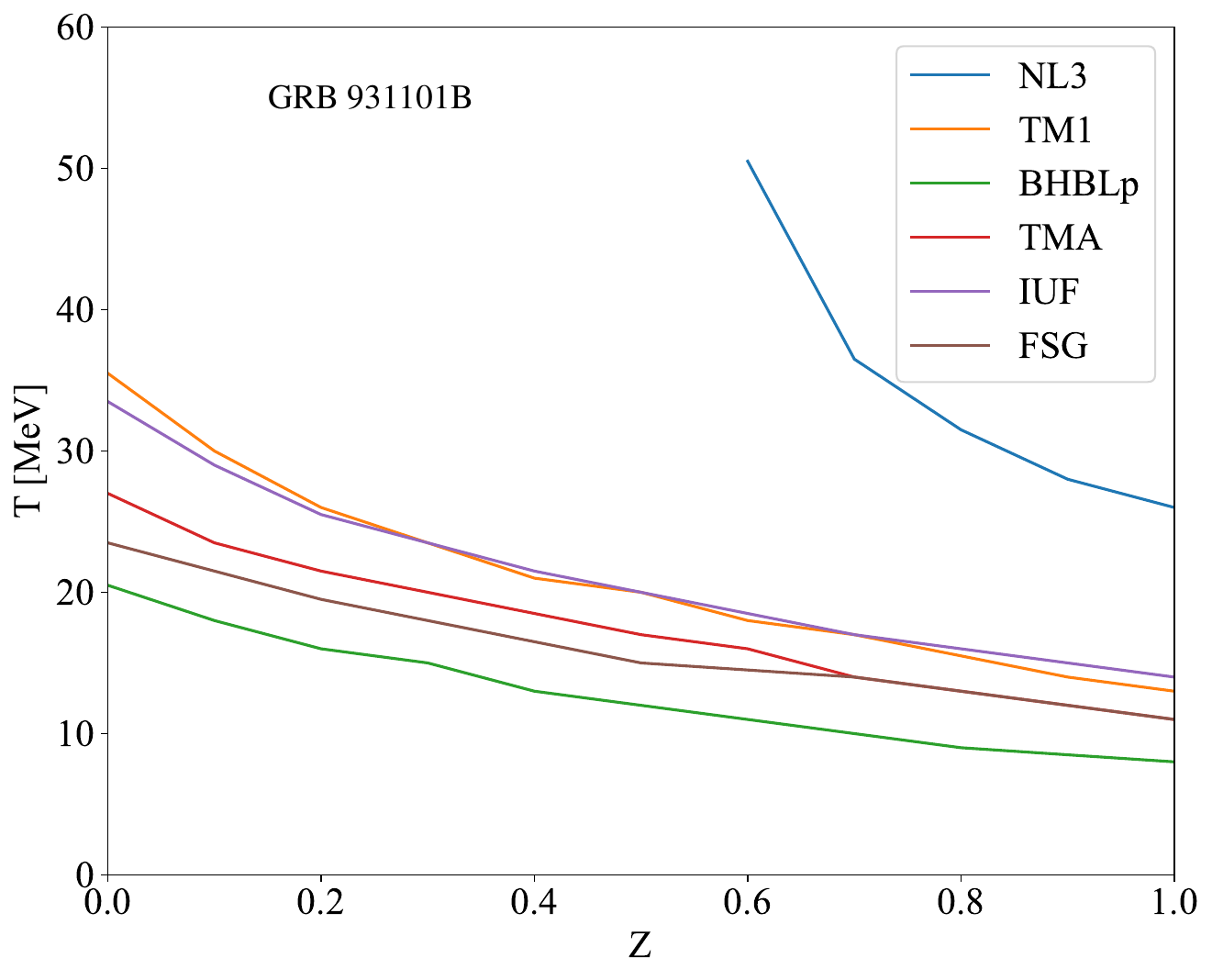}
            \includegraphics[width=0.49\textwidth]{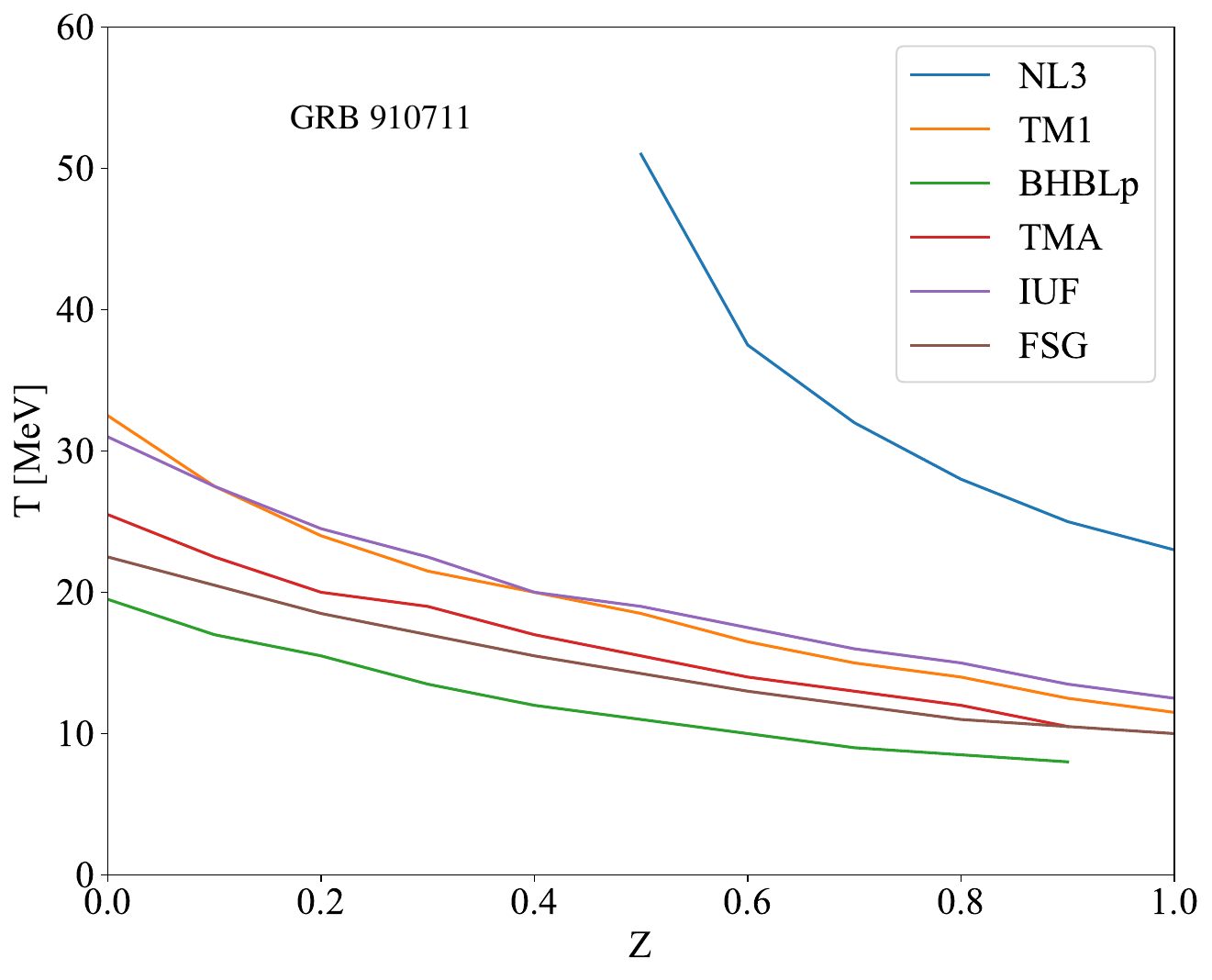}
            \caption{The relationship between the temperature $T$ and the redshift $z$ for different EOS models. The temperature values correspond to the points where the $\text{Metric}<0$ (from Figure \ref{Fig5}) and reaches its minimum, ensuring overlap between the $\rho_c$ ranges of the $n=0$ and $n=1$ modes.}
            \label{Fig6}
        \end{figure*}

\begin{figure*}
            \centering
            \includegraphics[width=0.49\textwidth]{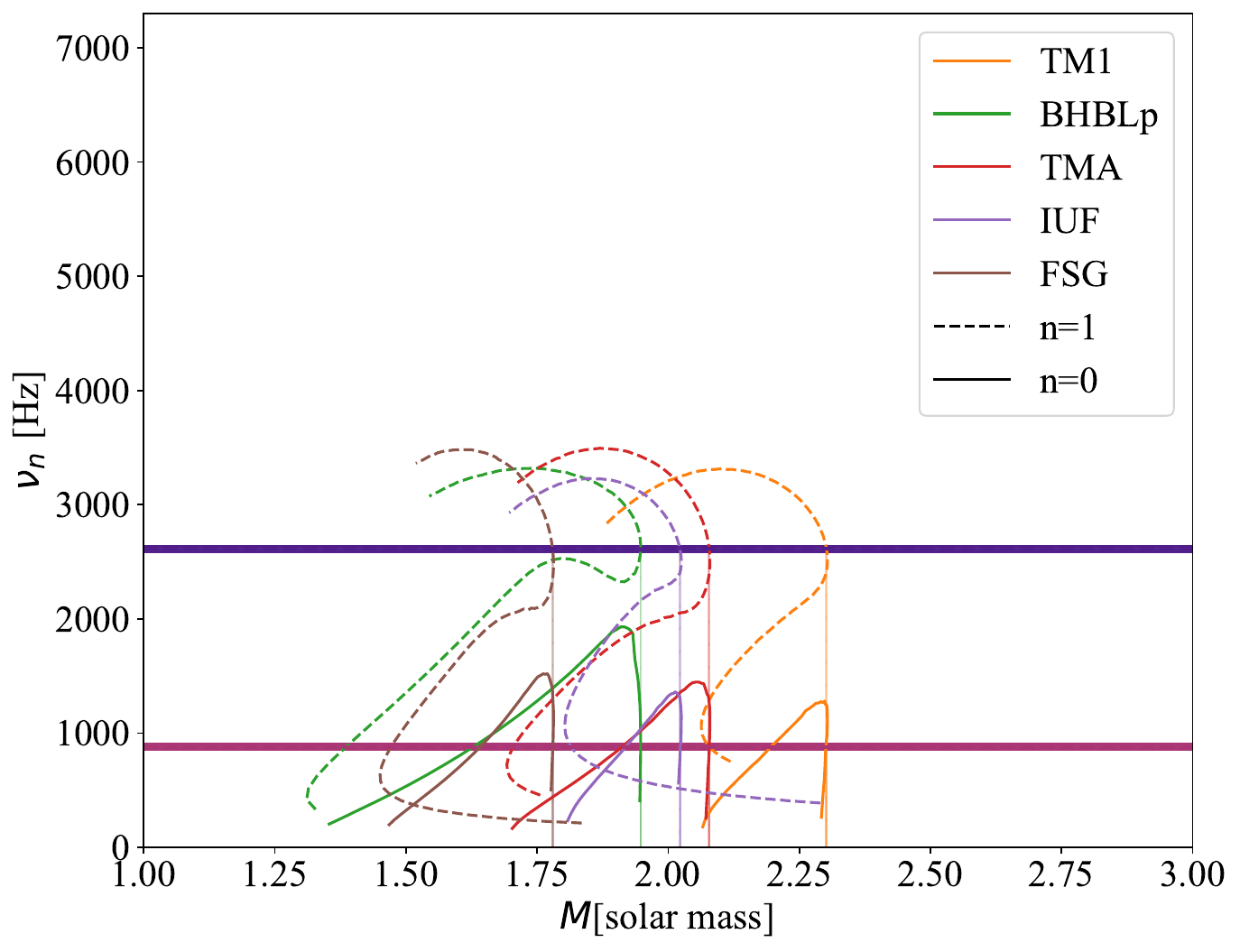}
            \includegraphics[width=0.49\textwidth]{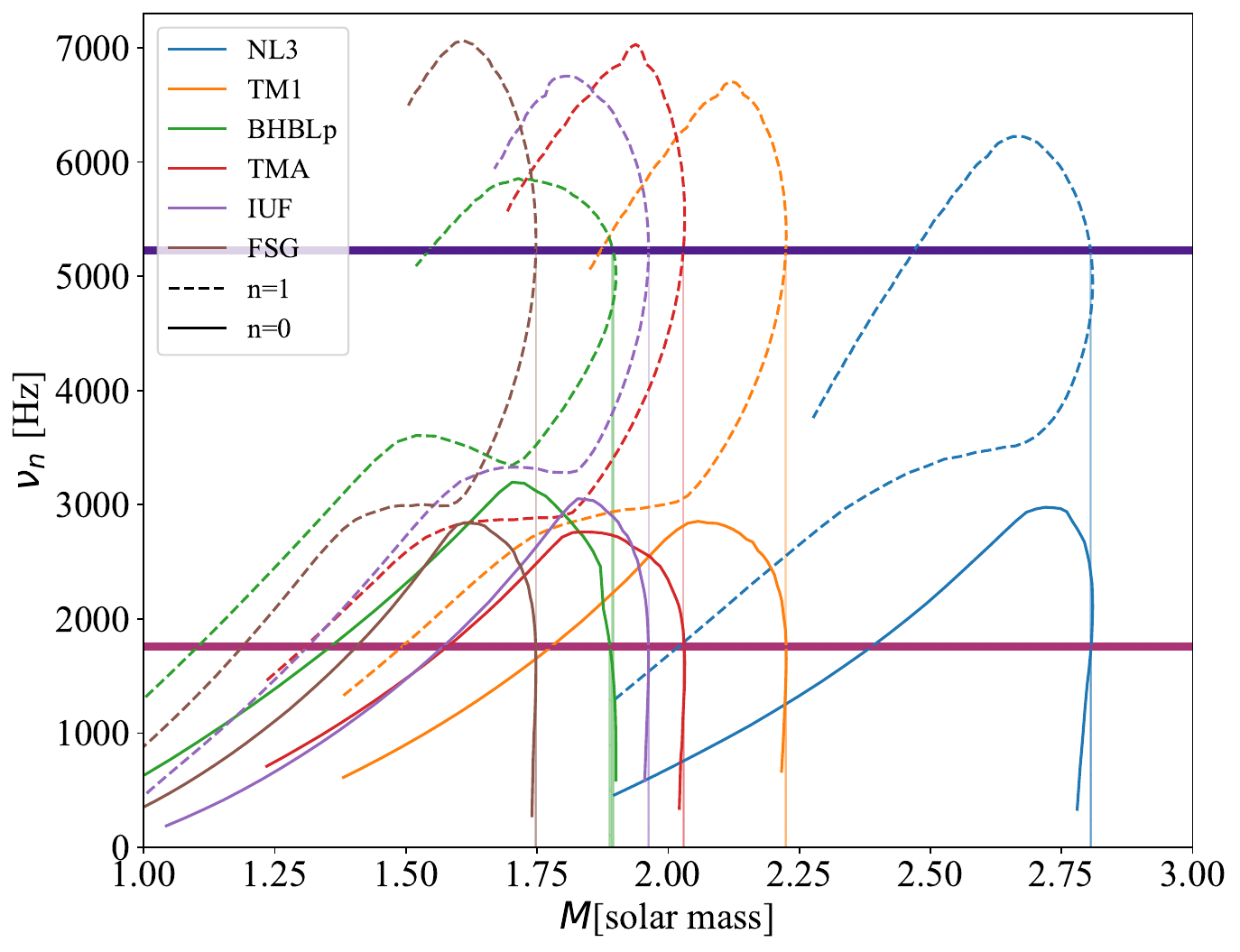}
            \caption{Eigenfrequencies $\nu_n$ of the $n=0$ (solid lines) and $n=1$ (dashed lines) modes as functions of NS mass for different EOS models. The left and right panels correspond to a redshift of $z=0$ and $z=1$, respectively. The temperatures corresponding to the minimum Metric values at $z=0$ and $z=1$ for each EOS are taken from Figure \ref{Fig6}. Horizontal lines are the observed QPO frequencies for GRB 931101B. The vertical shaded mass ranges highlight regions where the frequencies of both modes align with the observed QPOs.}
            \label{Fig7}
        \end{figure*}

        On the other hand, the effect of redshift in the above calculation is ignored due to the lack of redshift measurements. In fact, one needs to consider the uncertainty caused by the redshift of these two short GRBs. Figure \ref{Fig6} shows the relationship between the temperature \(T\) and the redshift \(z\) for different EOSs. Here temperature \(T\) values correspond to the points where the Metric reaches its minimum value. For example, the minimum value of temperature in Figure \ref{Fig5} for a given EOS corresponds to $z=0$. Then, we perform the calculations by varying the redshift range from 0 to 1 based on the observations.\footnote{\cite{L"u.at.al.2015} found that the redshift distribution for short GRBs with redshift measured is from 0 to 1, and average redshift is about $z=0.58$.} We find that the temperature decreases as the redshift increases, and it is due to the relativistic effect of redshift that increases the frequency in the rest frame. Moreover, it is found that the NL3 EOS can reach the overlap of \(\rho_c\) ranges at a higher temperature compared to other EOSs when $z\gtrsim0.5$. For example, Figure \ref{Fig7} shows the eigenfrequency as a function of the NS mass for $n=0$ (solid lines) and $n=1$ (dashed lines) oscillation modes for different EOSs at $z=0$ and $z=1$, respectively.

        On the other hand, the effect of redshift in the above calculation is ignored due to non-redshift measured. In fact, one needs to consider the uncertainty caused by the redshift of these two short GRBs. Fig.~\ref{Fig6} shows the relationship between the temperature \(T\) and the redshift \(z\) for different EOSs. Here, temperature \(T\) values are corresponding to the points where the Metric reaches its minimum value. For example, the minimum value of temperature in Fig.~\ref{Fig5} for a given EOS is corresponding to $z=0$. Then, we perform the calculations by varying the redshift range from 0 to 1 based on the observations\footnote{\cite{L"u.at.al.2015} found that the redshift distribution for short GRBs with redshift measured is from 0 to 1, and the average redshift is about $z=0.58$.}. We find that the temperature decreases as the redshift increases, and it is due to the relativistic effect of redshift which increases the frequency in the rest frame. Moreover, it is found that the NL3 EOS can reach the overlap of \(\rho_c\) ranges at a higher temperature compared to other EOSs when $z\gtrsim0.5$. For example, Fig.~\ref{Fig7} shows the eigen-frequency as a function of the NS mass for $n=0$ (solid lines) and $n=1$ (dashed lines) oscillation modes for different EOSs at $z=0$ and $z=1$, respectively.

        In any case, it is difficult to constrain the EOS of NSs based solely on the observed high-frequency QPO in those two short GRBs, and we cannot rule out any EOS by considering the effect from both temperature and redshift. However, it is found that the different EOSs of NSs correspond to different temperatures for a given redshift with $\text{Metric}=0$. In other words, if we believe that the observed high-frequency QPOs in short GRBs are indeed caused by radial oscillations of HMNSs from binary NS mergers, one can constrain the EOS only when the redshift and temperature of the remnant can be measured.

        \section{Conclusion and Discussion}
        NSs offer unparalleled opportunities for investigating extreme physics phenomena in the universe. Gaining insights into the composition of matter and the structure of NSs would be a significant advancement in nuclear physics and astrophysics. The application of seismology in QPOs of SGRs has demonstrated immense potential in this regard, particularly in providing constraints on the EOS and the thickness of the crust of NSs. The recent identification of QPOs from two short-duration GRBs (GRB 931101B and GRB 910711) may provide a new opportunity to explore the nature of NSs by employing seismology.
        
        In this paper, we investigate the potential of torsional and radial oscillation models to explain the observed high-frequency QPOs in two short-duration GRBs (GRB 931101B and GRB 910711), and then we try to constrain the EOS of NSs. The following interesting results can be obtained:
        \begin{itemize}	
        \item[$1.$] \textit{Torsional oscillation}. By assuming that the observed short-duration GRBs (GRB 931103B and GRB 910711) with high-frequency QPOs originated from GFs of magnetars with torsional oscillation, the selected six EOSs (i.e., TM1, NL3, APR, SLy4, DDME2, and GM1) of NSs that are suitable for the condition of zero temperature (i.e., cold crust) exhibit significant overlapping in mass ranges, and this suggests that we cannot constrain the EOS of NSs via the high-frequency QPOs in those two short-duration GBRs.
        \item[$2.$] \textit{Radial oscillation}. By assuming that the observed shortduration GRBs (GRB 931103B and GRB 910711) with high-frequency QPOs originated from the radial oscillation of an HMNS from a binary NS merger, the selected six EOSs (i.e., IUF, TM1, TMA, FSG, BHBLp, and NL3) of NSs that are suitable for the condition of high temperature (i.e., hot remnant) cannot be ruled out with redshift considered. However, it is found that the different EOSs of NSs correspond to different temperatures for a given redshift with Metric = 0. This means that the EOS can only be constrained if the redshift and temperature of the remnant can be measured.
        \end{itemize}
        
        On the other hand, the origin of these two short GRBs may be related to the inspiral phase of the binary NS merger. This phase could also explain the observed high-frequency QPOs, which caused by torsional oscillation. If this is the case, the $\gamma$-ray emission of short GRBs could result from the interaction of the NS magnetospheres during the inspiral phase \citep{Vietri.1996, Hansen.Lyutikov.2001, McWilliams.Levin.2011, Palenzuela.et.al.2013, Wang.et.al.2018}, or from the cracking of the tidal crust in the last moments of a compact binary merger \citep{Troja.et.al.2010}. In this framework, the GRB could be identified as an isolated precursor rather than being associated with relativistic outflows from post-merger remnants \citep{Burlon.et.al.2008,Zhong.et.al.2019,Coppin.et.al.2020,Wang.et.al.2020}. Unlike the fluid-dominated, high-temperature environment of an HMNS formed post-merger, the inspiral phase involves relatively cold NSs with intact crusts, where shear stresses can excite torsional modes. The maximum luminosity of the precursor from the interaction of NS magnetospheres is proportional to $B^2$ . In fact, this requires introducing an additional assumption, namely, the binary system must contain at least one NS with a magnetar-like magnetic field ($B > 10^{15}$ G) to achieve the observed luminosity \citep{Wang.et.al.2020}.

        Recently, numerical relativity studies have suggested that the quasi-periodic kHz substructure observed in electromagnetic outflows from HMNSs may be dominated by magnetohydro-dynamic (MHD) shearing processes rather than being related to internal stellar oscillations \citep{Most.Quataert.2023}. This indicates that QPOs in relativistic outflows may not be necessarily related to internal NS oscillations, making it difficult to constrain the EOS of NSs. This is consistent with the results obtained in this work. However, asteroseismology remains a potential method to constrain the EOS of NSs when more precise information of the NS temperature can be obtained, such as the redshift and magnetic field strength. Moreover, \cite{Guedes.et.al.2024} adopted numerical relativity simulations of post-merger remnants and their relationship to constrain the binary tidal coupling constant, and then they constrain the EOS of NSs by invoking the observed highfrequency QPOs in those two short GRBs. They claim to have found tight constraints on the mass–radius relation, which differs significantly from what we have obtained in this paper. The possible reason is that the adopted method and the physical conditions are different from each other. Our method can constrain the redshift based on the presence of two distinct QPO frequencies ($\nu_1$ and $\nu_2$) in these bursts, with certain strong assumptions (e.g., temperature). In contrast, the approach from \cite{Guedes.et.al.2024} adopted $\nu_2$ and the ratio $\nu_2/\nu_1$ to constrain the redshift, and the ratio between $\nu_2$ and $\nu_1$ is redshift independent. We hope that multimessenger observations, combining electromagnetic signals and GWs, will offer more precise constraints on the physical properties of NS remnants in the future, such as their temperature, redshift, and magnetic field strength.
        
        \begin{acknowledgements}
        This work is supported by the Guangxi Science Foundation (grant No. 2023GXNSFDA026007), the National Natural Science Foundation of China (grant Nos. 12494574, 11922301, and 12133003), the Program of Bagui Scholars Program (LHJ), and the Guangxi Talent Program (“Highland of Innovation Talents”).
        \end{acknowledgements}

\bibliography{ms}{}

\begin{thebibliography}{}
\expandafter\ifx\csname natexlab\endcsname\relax\def\natexlab#1{#1}\fi
\providecommand{\url}[1]{\href{#1}{#1}}
\providecommand{\dodoi}[1]{doi:~\href{http://doi.org/#1}{\nolinkurl{#1}}}
\providecommand{\doeprint}[1]{\href{http://ascl.net/#1}{\nolinkurl{http://ascl.net/#1}}}
\providecommand{\doarXiv}[1]{\href{https://arxiv.org/abs/#1}{\nolinkurl{https://arxiv.org/abs/#1}}}

\bibitem[{{Abbott} {et~al.}(2017){Abbott}, {Abbott}, {Abbott}, {Acernese}, {Ackley}, {Adams}, {Adams}, {Addesso}, {Adhikari}, {Adya}, {Affeldt}, {Afrough}, {Agarwal}, {Agathos}, {Agatsuma}, {Aggarwal}, {Aguiar}, \& {Aiello}}]{Abbott.et.al.2017}
{Abbott}, B.~P., {Abbott}, R., {Abbott}, T.~D., {et~al.} 2017, \prl, 119, 161101, \dodoi{10.1103/PhysRevLett.119.161101}

\bibitem[{{Akmal} {et~al.}(1998){Akmal}, {Pandharipande}, \& {Ravenhall}}]{Akmal.et.al.1998}
{Akmal}, A., {Pandharipande}, V.~R., \& {Ravenhall}, D.~G. 1998, \prc, 58, 1804, \dodoi{10.1103/PhysRevC.58.1804}

\bibitem[{{Anderson} {et~al.}(1990){Anderson}, {Gorham}, {Kulkarni}, {Prince}, \& {Wolszczan}}]{Anderson.et.al.1990}
{Anderson}, S.~B., {Gorham}, P.~W., {Kulkarni}, S.~R., {Prince}, T.~A., \& {Wolszczan}, A. 1990, \nat, 346, 42, \dodoi{10.1038/346042a0}

\bibitem[{{Bauswein} {et~al.}(2010){Bauswein}, {Janka}, \& {Oechslin}}]{Bauswein.et.al.2010}
{Bauswein}, A., {Janka}, H.~T., \& {Oechslin}, R. 2010, \prd, 82, 084043, \dodoi{10.1103/PhysRevD.82.084043}

\bibitem[{{Bauswein} {et~al.}(2020){Bauswein}, {Blacker}, {Vijayan}, {Stergioulas}, {Chatziioannou}, {Clark}, {Bastian}, {Blaschke}, {Cierniak}, \& {Fischer}}]{Bauswein.et.al.2020}
{Bauswein}, A., {Blacker}, S., {Vijayan}, V., {et~al.} 2020, \prl, 125, 141103, \dodoi{10.1103/PhysRevLett.125.141103}

\bibitem[{{Baym} {et~al.}(1971){Baym}, {Pethick}, \& {Sutherland}}]{BPS.1971}
{Baym}, G., {Pethick}, C., \& {Sutherland}, P. 1971, \apj, 170, 299, \dodoi{10.1086/151216}

\bibitem[{{Burgay} {et~al.}(2003){Burgay}, {D'Amico}, {Possenti}, {Manchester}, {Lyne}, {Joshi}, {McLaughlin}, {Kramer}, {Sarkissian}, {Camilo}, {Kalogera}, {Kim}, \& {Lorimer}}]{Burgay.et.al.2003}
{Burgay}, M., {D'Amico}, N., {Possenti}, A., {et~al.} 2003, \nat, 426, 531, \dodoi{10.1038/nature02124}

\bibitem[{{Burlon} {et~al.}(2008){Burlon}, {Ghirlanda}, {Ghisellini}, {Lazzati}, {Nava}, {Nardini}, \& {Celotti}}]{Burlon.et.al.2008}
{Burlon}, D., {Ghirlanda}, G., {Ghisellini}, G., {et~al.} 2008, \apjl, 685, L19, \dodoi{10.1086/592350}

\bibitem[{{Burns} {et~al.}(2021){Burns}, {Svinkin}, {Hurley}, {Wadiasingh}, {Negro}, {Younes}, {Hamburg}, {Ridnaia}, {Cook}, {Cenko}, {Aloisi}, {Ashton}, {Baring}, {Briggs}, {Christensen}, {Frederiks}, {Goldstein}, {Hui}, {Kaplan}, {Kasliwal}, {Kocevski}, {Roberts}, {Savchenko}, {Tohuvavohu}, {Veres}, \& {Wilson-Hodge}}]{Burns.et.al.2021}
{Burns}, E., {Svinkin}, D., {Hurley}, K., {et~al.} 2021, \apjl, 907, L28, \dodoi{10.3847/2041-8213/abd8c8}

\bibitem[{{Chandrasekhar}(1964)}]{S.Chandrasekhar.1964}
{Chandrasekhar}, S. 1964, \apj, 140, 417, \dodoi{10.1086/147938}

\bibitem[{{Chanmugam}(1977)}]{G.Chanmugam.1977}
{Chanmugam}, G. 1977, \apj, 217, 799, \dodoi{10.1086/155627}

\bibitem[{{Chirenti} {et~al.}(2023){Chirenti}, {Dichiara}, {Lien}, {Miller}, \& {Preece}}]{Chirenti.et.al.2023}
{Chirenti}, C., {Dichiara}, S., {Lien}, A., {Miller}, M.~C., \& {Preece}, R. 2023, \nat, 613, 253, \dodoi{10.1038/s41586-022-05497-0}

\bibitem[{{Coppin} {et~al.}(2020){Coppin}, {de Vries}, \& {van Eijndhoven}}]{Coppin.et.al.2020}
{Coppin}, P., {de Vries}, K.~D., \& {van Eijndhoven}, N. 2020, \prd, 102, 103014, \dodoi{10.1103/PhysRevD.102.103014}

\bibitem[{{Dai} \& {Lu}(1998{\natexlab{a}})}]{Dai.and.lu.1998a}
{Dai}, Z.~G., \& {Lu}, T. 1998{\natexlab{a}}, \prl, 81, 4301, \dodoi{10.1103/PhysRevLett.81.4301}

\bibitem[{{Dai} \& {Lu}(1998{\natexlab{b}})}]{Dai.and.lu.1998b}
---. 1998{\natexlab{b}}, \aap, 333, L87, \dodoi{10.48550/arXiv.astro-ph/9810402}

\bibitem[{{Dai} {et~al.}(2006){Dai}, {Wang}, {Wu}, \& {Zhang}}]{Dai.et.al.2006}
{Dai}, Z.~G., {Wang}, X.~Y., {Wu}, X.~F., \& {Zhang}, B. 2006, Science, 311, 1127, \dodoi{10.1126/science.1123606}

\bibitem[{{Douchin} \& {Haensel}(2000)}]{Douchin.Haensel.2000}
{Douchin}, F., \& {Haensel}, P. 2000, Physics Letters B, 485, 107, \dodoi{10.1016/S0370-2693(00)00672-9}

\bibitem[{{Douchin} \& {Haensel}(2001)}]{Douchin.Haensel.2001}
---. 2001, \aap, 380, 151, \dodoi{10.1051/0004-6361:20011402}

\bibitem[{{Duncan}(1998)}]{Duncan.1998}
{Duncan}, R.~C. 1998, \apjl, 498, L45, \dodoi{10.1086/311303}

\bibitem[{{Duncan} \& {Thompson}(1992)}]{Duncan.Thompson.1992}
{Duncan}, R.~C., \& {Thompson}, C. 1992, \apjl, 392, L9, \dodoi{10.1086/186413}

\bibitem[{{Fan} \& {Xu}(2006)}]{Fan.Xu.2006}
{Fan}, Y.-Z., \& {Xu}, D. 2006, \mnras, 372, L19, \dodoi{10.1111/j.1745-3933.2006.00217.x}

\bibitem[{{Fonseca} {et~al.}(2021){Fonseca}, {Cromartie}, {Pennucci}, {Ray}, {Kirichenko}, {Ransom}, {Demorest}, {Stairs}, {Arzoumanian}, {Guillemot}, {Parthasarathy}, {Kerr}, {Cognard}, {Baker}, {Blumer}, {Brook}, {DeCesar}, {Dolch}, {Dong}, {Ferrara}, {Fiore}, {Garver-Daniels}, {Good}, {Jennings}, {Jones}, {Kaspi}, {Lam}, {Lorimer}, {Luo}, {McEwen}, {McKee}, {McLaughlin}, {McMann}, {Meyers}, {Naidu}, {Ng}, {Nice}, {Pol}, {Radovan}, {Shapiro-Albert}, {Tan}, {Tendulkar}, {Swiggum}, {Wahl}, \& {Zhu}}]{Fonseca.et.al.2021}
{Fonseca}, E., {Cromartie}, H.~T., {Pennucci}, T.~T., {et~al.} 2021, \apjl, 915, L12, \dodoi{10.3847/2041-8213/ac03b8}

\bibitem[{{Gabler} {et~al.}(2011){Gabler}, {Cerd{\'a} Dur{\'a}n}, {Font}, {M{\"u}ller}, \& {Stergioulas}}]{Gabler.et.al.2011}
{Gabler}, M., {Cerd{\'a} Dur{\'a}n}, P., {Font}, J.~A., {M{\"u}ller}, E., \& {Stergioulas}, N. 2011, \mnras, 410, L37, \dodoi{10.1111/j.1745-3933.2010.00974.x}

\bibitem[{{Gabler} {et~al.}(2012){Gabler}, {Cerd{\'a}-Dur{\'a}n}, {Stergioulas}, {Font}, \& {M{\"u}ller}}]{Gabler.et.al.2012}
{Gabler}, M., {Cerd{\'a}-Dur{\'a}n}, P., {Stergioulas}, N., {Font}, J.~A., \& {M{\"u}ller}, E. 2012, \mnras, 421, 2054, \dodoi{10.1111/j.1365-2966.2012.20454.x}

\bibitem[{{Gao} {et~al.}(2017){Gao}, {Zhang}, {L{\"u}}, \& {Li}}]{Gao.et.al.2017}
{Gao}, H., {Zhang}, B., {L{\"u}}, H.-J., \& {Li}, Y. 2017, \apj, 837, 50, \dodoi{10.3847/1538-4357/aa5be3}

\bibitem[{{Giacomazzo} \& {Perna}(2013)}]{Giacomazzo.Perna.2013}
{Giacomazzo}, B., \& {Perna}, R. 2013, \apjl, 771, L26, \dodoi{10.1088/2041-8205/771/2/L26}

\bibitem[{Glendenning \& Moszkowski(1991)}]{Glendenning.Moszkowski.1991}
Glendenning, N.~K., \& Moszkowski, S.~A. 1991, Phys. Rev. Lett., 67, 2414, \dodoi{10.1103/PhysRevLett.67.2414}

\bibitem[{{Goldstein} {et~al.}(2017){Goldstein}, {Veres}, {Burns}, {Briggs}, {Hamburg}, {Kocevski}, {Wilson-Hodge}, {Preece}, {Poolakkil}, {Roberts}, {Hui}, {Connaughton}, {Racusin}, {von Kienlin}, {Dal Canton}, {Christensen}, {Littenberg}, {Siellez}, {Blackburn}, {Broida}, {Bissaldi}, {Cleveland}, {Gibby}, {Giles}, {Kippen}, {McBreen}, {McEnery}, {Meegan}, {Paciesas}, \& {Stanbro}}]{Goldstein.et.al.2017}
{Goldstein}, A., {Veres}, P., {Burns}, E., {et~al.} 2017, \apjl, 848, L14, \dodoi{10.3847/2041-8213/aa8f41}

\bibitem[{{Gondek} {et~al.}(1997{\natexlab{a}}){Gondek}, {Haensel}, \& {Zdunik}}]{D.Gondek.1997}
{Gondek}, D., {Haensel}, P., \& {Zdunik}, J.~L. 1997{\natexlab{a}}, \aap, 325, 217, \dodoi{10.48550/arXiv.astro-ph/9705157}

\bibitem[{{Gondek} {et~al.}(1997{\natexlab{b}}){Gondek}, {Haensel}, \& {Zdunik}}]{Gondek.et.al.1997}
---. 1997{\natexlab{b}}, \aap, 325, 217, \dodoi{10.48550/arXiv.astro-ph/9705157}

\bibitem[{{Guedes} {et~al.}(2024){Guedes}, {Radice}, {Chirenti}, \& {Yagi}}]{Guedes.et.al.2024}
{Guedes}, V., {Radice}, D., {Chirenti}, C., \& {Yagi}, K. 2024, arXiv e-prints, arXiv:2408.16534, \dodoi{10.48550/arXiv.2408.16534}

\bibitem[{{Haensel} \& {Pichon}(1994)}]{Haensel.Pichon.1994}
{Haensel}, P., \& {Pichon}, B. 1994, \aap, 283, 313, \dodoi{10.48550/arXiv.nucl-th/9310003}

\bibitem[{{Haensel} {et~al.}(2007){Haensel}, {Potekhin}, \& {Yakovlev}}]{Haensel.et.al.2007}
{Haensel}, P., {Potekhin}, A.~Y., \& {Yakovlev}, D.~G. 2007, {Neutron Stars 1 : Equation of State and Structure}, Vol. 326 (Springer New York, NY)

\bibitem[{{Hansen} \& {Lyutikov}(2001)}]{Hansen.Lyutikov.2001}
{Hansen}, B. M.~S., \& {Lyutikov}, M. 2001, \mnras, 322, 695, \dodoi{10.1046/j.1365-8711.2001.04103.x}

\bibitem[{{Hulse} \& {Taylor}(1975)}]{Hulse.Taylor.1975}
{Hulse}, R.~A., \& {Taylor}, J.~H. 1975, \apjl, 195, L51, \dodoi{10.1086/181708}

\bibitem[{Lalazissis {et~al.}(1997)Lalazissis, K\"onig, \& Ring}]{Lalazissis.et.al.1997}
Lalazissis, G.~A., K\"onig, J., \& Ring, P. 1997, Phys. Rev. C, 55, 540, \dodoi{10.1103/PhysRevC.55.540}

\bibitem[{Lalazissis {et~al.}(2005)Lalazissis, Nik\ifmmode \check{s}\else \v{s}\fi{}i\ifmmode~\acute{c}\else \'{c}\fi{}, Vretenar, \& Ring}]{Lalazissis.et.al.2005}
Lalazissis, G.~A., Nik\ifmmode \check{s}\else \v{s}\fi{}i\ifmmode~\acute{c}\else \'{c}\fi{}, T., Vretenar, D., \& Ring, P. 2005, Phys. Rev. C, 71, 024312, \dodoi{10.1103/PhysRevC.71.024312}

\bibitem[{{Lan} {et~al.}(2020){Lan}, {L{\"u}}, {Rice}, \& {Liang}}]{Lan.et.al.2020}
{Lan}, L., {L{\"u}}, H.-J., {Rice}, J., \& {Liang}, E.-W. 2020, \apj, 890, 99, \dodoi{10.3847/1538-4357/ab6c64}

\bibitem[{{Lasky} {et~al.}(2014){Lasky}, {Haskell}, {Ravi}, {Howell}, \& {Coward}}]{Lasky.et.al.2014}
{Lasky}, P.~D., {Haskell}, B., {Ravi}, V., {Howell}, E.~J., \& {Coward}, D.~M. 2014, \prd, 89, 047302, \dodoi{10.1103/PhysRevD.89.047302}

\bibitem[{{Lattimer}(2021)}]{Lattimer.2021}
{Lattimer}, J.~M. 2021, Annual Review of Nuclear and Particle Science, 71, 433, \dodoi{10.1146/annurev-nucl-102419-124827}

\bibitem[{{Lattimer} \& {Prakash}(2001)}]{Lattimer.Prakash.2001}
{Lattimer}, J.~M., \& {Prakash}, M. 2001, \apj, 550, 426, \dodoi{10.1086/319702}

\bibitem[{{Lee}(2007)}]{Lee.2007}
{Lee}, U. 2007, \mnras, 374, 1015, \dodoi{10.1111/j.1365-2966.2006.11214.x}

\bibitem[{{L{\"u}} {et~al.}(2015){L{\"u}}, {Zhang}, {Lei}, {Li}, \& {Lasky}}]{L"u.at.al.2015}
{L{\"u}}, H.-J., {Zhang}, B., {Lei}, W.-H., {Li}, Y., \& {Lasky}, P.~D. 2015, \apj, 805, 89, \dodoi{10.1088/0004-637X/805/2/89}

\bibitem[{{L{\"u}} {et~al.}(2017){L{\"u}}, {Zhang}, {Zhong}, {Hou}, {Sun}, {Rice}, \& {Liang}}]{Lv.et.al.2017}
{L{\"u}}, H.-J., {Zhang}, H.-M., {Zhong}, S.-Q., {et~al.} 2017, \apj, 835, 181, \dodoi{10.3847/1538-4357/835/2/181}

\bibitem[{{McDermott} {et~al.}(1988){McDermott}, {van Horn}, \& {Hansen}}]{McDermott.et.al.1988}
{McDermott}, P.~N., {van Horn}, H.~M., \& {Hansen}, C.~J. 1988, \apj, 325, 725, \dodoi{10.1086/166044}

\bibitem[{{McWilliams} \& {Levin}(2011)}]{McWilliams.Levin.2011}
{McWilliams}, S.~T., \& {Levin}, J. 2011, \apj, 742, 90, \dodoi{10.1088/0004-637X/742/2/90}

\bibitem[{{Messios} {et~al.}(2001){Messios}, {Papadopoulos}, \& {Stergioulas}}]{Messios.Papadopoulos.2001}
{Messios}, N., {Papadopoulos}, D.~B., \& {Stergioulas}, N. 2001, \mnras, 328, 1161, \dodoi{10.1046/j.1365-8711.2001.04645.x}

\bibitem[{{Most} \& {Quataert}(2023)}]{Most.Quataert.2023}
{Most}, E.~R., \& {Quataert}, E. 2023, \apjl, 947, L15, \dodoi{10.3847/2041-8213/acca84}

\bibitem[{{Negele} \& {Vautherin}(1973)}]{Negele.Vautherin.1973}
{Negele}, J.~W., \& {Vautherin}, D. 1973, \nphysa, 207, 298, \dodoi{10.1016/0375-9474(73)90349-7}

\bibitem[{{Oechslin} {et~al.}(2007){Oechslin}, {Janka}, \& {Marek}}]{Oechslin.et.al.2007}
{Oechslin}, R., {Janka}, H.~T., \& {Marek}, A. 2007, \aap, 467, 395, \dodoi{10.1051/0004-6361:20066682}

\bibitem[{Oertel {et~al.}(2017)Oertel, Hempel, Klaehn, \& Typel}]{Oertel:2016bki}
Oertel, M., Hempel, M., Klaehn, T., \& Typel, S. 2017, Rev. Mod. Phys., 89, 015007, \dodoi{10.1103/RevModPhys.89.015007}

\bibitem[{Ogata \& Ichimaru(1990)}]{Ogata.Ichimaru.1990}
Ogata, S., \& Ichimaru, S. 1990, Phys. Rev. A, 42, 4867, \dodoi{10.1103/PhysRevA.42.4867}

\bibitem[{{Oppenheimer} \& {Volkoff}(1939)}]{Oppenheimer.Volkoff.1939}
{Oppenheimer}, J.~R., \& {Volkoff}, G.~M. 1939, Physical Review, 55, 374, \dodoi{10.1103/PhysRev.55.374}

\bibitem[{{Palenzuela} {et~al.}(2013){Palenzuela}, {Lehner}, {Ponce}, {Liebling}, {Anderson}, {Neilsen}, \& {Motl}}]{Palenzuela.et.al.2013}
{Palenzuela}, C., {Lehner}, L., {Ponce}, M., {et~al.} 2013, \prl, 111, 061105, \dodoi{10.1103/PhysRevLett.111.061105}

\bibitem[{{Piro}(2005)}]{Piro.2005}
{Piro}, A.~L. 2005, \apjl, 634, L153, \dodoi{10.1086/499049}

\bibitem[{{Rezzolla} {et~al.}(2010){Rezzolla}, {Baiotti}, {Giacomazzo}, {Link}, \& {Font}}]{Rezzolla.et.al.2010}
{Rezzolla}, L., {Baiotti}, L., {Giacomazzo}, B., {Link}, D., \& {Font}, J.~A. 2010, Classical and Quantum Gravity, 27, 114105, \dodoi{10.1088/0264-9381/27/11/114105}

\bibitem[{{Roberts} {et~al.}(2021){Roberts}, {Veres}, {Baring}, {Briggs}, {Kouveliotou}, {Bissaldi}, {Younes}, {Chastain}, {DeLaunay}, {Huppenkothen}, {Tohuvavohu}, {Bhat}, {G{\"o}{\v{g}}{\"u}{\c{s}}}, {van der Horst}, {Kennea}, {Kocevski}, {Linford}, {Guiriec}, {Hamburg}, {Wilson-Hodge}, \& {Burns}}]{Roberts.et.al.2021}
{Roberts}, O.~J., {Veres}, P., {Baring}, M.~G., {et~al.} 2021, \nat, 589, 207, \dodoi{10.1038/s41586-020-03077-8}

\bibitem[{{Rosswog} {et~al.}(2000){Rosswog}, {Davies}, {Thielemann}, \& {Piran}}]{Rosswog.2000}
{Rosswog}, S., {Davies}, M.~B., {Thielemann}, F.~K., \& {Piran}, T. 2000, \aap, 360, 171, \dodoi{10.48550/arXiv.astro-ph/0005550}

\bibitem[{{Sagun} {et~al.}(2020){Sagun}, {Panotopoulos}, \& {Lopes}}]{Sagun.2020}
{Sagun}, V., {Panotopoulos}, G., \& {Lopes}, I. 2020, \prd, 101, 063025, \dodoi{10.1103/PhysRevD.101.063025}

\bibitem[{{Samuelsson} \& {Andersson}(2007)}]{Samuelsson.Andersson.2007}
{Samuelsson}, L., \& {Andersson}, N. 2007, \mnras, 374, 256, \dodoi{10.1111/j.1365-2966.2006.11147.x}

\bibitem[{{Schumaker} \& {Thorne}(1983)}]{Schumaker.Thorne.1983}
{Schumaker}, B.~L., \& {Thorne}, K.~S. 1983, \mnras, 203, 457, \dodoi{10.1093/mnras/203.2.457}

\bibitem[{{Shapiro} \& {Teukolsky}(1983)}]{Shapiro.et.al.1983}
{Shapiro}, S.~L., \& {Teukolsky}, S.~A. 1983, {Black holes, white dwarfs and neutron stars. The physics of compact objects} (Wiley), \dodoi{10.1002/9783527617661}

\bibitem[{{Shibata}(2005)}]{Shibata.2005}
{Shibata}, M. 2005, \prl, 94, 201101, \dodoi{10.1103/PhysRevLett.94.201101}

\bibitem[{{Sotani} {et~al.}(2007){Sotani}, {Kokkotas}, \& {Stergioulas}}]{Sotani.et.al.2007}
{Sotani}, H., {Kokkotas}, K.~D., \& {Stergioulas}, N. 2007, \mnras, 375, 261, \dodoi{10.1111/j.1365-2966.2006.11304.x}

\bibitem[{Sotani {et~al.}(2012)Sotani, Nakazato, Iida, \& Oyamatsu}]{Sotani.et.al.2012}
Sotani, H., Nakazato, K., Iida, K., \& Oyamatsu, K. 2012, Phys. Rev. Lett., 108, 201101, \dodoi{10.1103/PhysRevLett.108.201101}

\bibitem[{{Steiner} \& {Watts}(2009)}]{Steiner.et.al.2009}
{Steiner}, A.~W., \& {Watts}, A.~L. 2009, \prl, 103, 181101, \dodoi{10.1103/PhysRevLett.103.181101}

\bibitem[{{Stergioulas}(2003)}]{Stergioulas.2003}
{Stergioulas}, N. 2003, Living Reviews in Relativity, 6, 3, \dodoi{10.12942/lrr-2003-3}

\bibitem[{{Stergioulas} {et~al.}(2011){Stergioulas}, {Bauswein}, {Zagkouris}, \& {Janka}}]{Stergioulas.et.al.2011}
{Stergioulas}, N., {Bauswein}, A., {Zagkouris}, K., \& {Janka}, H.-T. 2011, \mnras, 418, 427, \dodoi{10.1111/j.1365-2966.2011.19493.x}

\bibitem[{{Sugahara} \& {Toki}(1994)}]{Sugahara.Toki.1994}
{Sugahara}, Y., \& {Toki}, H. 1994, \nphysa, 579, 557, \dodoi{10.1016/0375-9474(94)90923-7}

\bibitem[{{Svinkin} {et~al.}(2021){Svinkin}, {Frederiks}, {Hurley}, {Aptekar}, {Golenetskii}, {Lysenko}, {Ridnaia}, {Tsvetkova}, {Ulanov}, {Cline}, {Mitrofanov}, {Golovin}, {Kozyrev}, {Litvak}, {Sanin}, {Goldstein}, {Briggs}, {Wilson-Hodge}, {von Kienlin}, {Zhang}, {Rau}, {Savchenko}, {Bozzo}, {Ferrigno}, {Ubertini}, {Bazzano}, {Rodi}, {Barthelmy}, {Cummings}, {Krimm}, {Palmer}, {Boynton}, {Fellows}, {Harshman}, {Enos}, \& {Starr}}]{Svinkin.et.al.2021}
{Svinkin}, D., {Frederiks}, D., {Hurley}, K., {et~al.} 2021, \nat, 589, 211, \dodoi{10.1038/s41586-020-03076-9}

\bibitem[{{Takami} {et~al.}(2014){Takami}, {Rezzolla}, \& {Baiotti}}]{Takami.et.al.2014}
{Takami}, K., {Rezzolla}, L., \& {Baiotti}, L. 2014, \prl, 113, 091104, \dodoi{10.1103/PhysRevLett.113.091104}

\bibitem[{{Thorne} \& {Campolattaro}(1967)}]{Thorne.Campolattaro.1967}
{Thorne}, K.~S., \& {Campolattaro}, A. 1967, {Non-Radial Pulsation of General-Relativistic Stellar Models. I. Analytic Analysis for L $\geq$ 2}, Astrophysical Journal, vol. 149, p.591, \dodoi{10.1086/149288}

\bibitem[{{Troja} {et~al.}(2010){Troja}, {Rosswog}, \& {Gehrels}}]{Troja.et.al.2010}
{Troja}, E., {Rosswog}, S., \& {Gehrels}, N. 2010, \apj, 723, 1711, \dodoi{10.1088/0004-637X/723/2/1711}

\bibitem[{Typel {et~al.}(2015)Typel, Oertel, \& Klaehn}]{Typel:2013rza}
Typel, S., Oertel, M., \& Klaehn, T. 2015, Phys. Part. Nucl., 46, 633, \dodoi{10.1134/S1063779615040061}

\bibitem[{{Vietri}(1996)}]{Vietri.1996}
{Vietri}, M. 1996, \apjl, 471, L95, \dodoi{10.1086/310340}

\bibitem[{{Wang} {et~al.}(2018){Wang}, {Peng}, {Wu}, \& {Dai}}]{Wang.et.al.2018}
{Wang}, J.-S., {Peng}, F.-K., {Wu}, K., \& {Dai}, Z.-G. 2018, \apj, 868, 19, \dodoi{10.3847/1538-4357/aae531}

\bibitem[{{Wang} {et~al.}(2020){Wang}, {Peng}, {Zou}, {Zhang}, \& {Zhang}}]{Wang.et.al.2020}
{Wang}, J.-S., {Peng}, Z.-K., {Zou}, J.-H., {Zhang}, B.-B., \& {Zhang}, B. 2020, \apjl, 902, L42, \dodoi{10.3847/2041-8213/abbfb8}

\bibitem[{{Watts} \& {Strohmayer}(2007)}]{Watts.Strohmayer.2007}
{Watts}, A.~L., \& {Strohmayer}, T.~E. 2007, Advances in Space Research, 40, 1446, \dodoi{10.1016/j.asr.2006.12.021}

\bibitem[{{Wolszczan}(1991)}]{Wolszczan.1991}
{Wolszczan}, A. 1991, \nat, 350, 688, \dodoi{10.1038/350688a0}

\bibitem[{{Yang} {et~al.}(2020){Yang}, {Chand}, {Zhang}, {Yang}, {Zou}, {Yang}, {Zhao}, {Shao}, {Xiong}, {Luo}, {Li}, {Xiao}, {Li}, {Liu}, {Joshi}, {Sharma}, {Chakraborty}, {Li}, \& {Zhang}}]{Yang.et.al.2020}
{Yang}, J., {Chand}, V., {Zhang}, B.-B., {et~al.} 2020, \apj, 899, 106, \dodoi{10.3847/1538-4357/aba745}

\bibitem[{{Zhang}(2013)}]{Zhang.2013}
{Zhang}, B. 2013, \apjl, 763, L22, \dodoi{10.1088/2041-8205/763/1/L22}

\bibitem[{{Zhang} {et~al.}(2018){Zhang}, {Zhang}, {Sun}, {Lei}, {Gao}, {Li}, {Shao}, {Zhao}, {Hu}, {L{\"u}}, {Wu}, {Fan}, {Wang}, {Castro-Tirado}, {Zhang}, {Yu}, {Cao}, \& {Liang}}]{Zhang.et.al.2018}
{Zhang}, B.~B., {Zhang}, B., {Sun}, H., {et~al.} 2018, Nature Communications, 9, 447, \dodoi{10.1038/s41467-018-02847-3}

\bibitem[{{Zhong} {et~al.}(2019){Zhong}, {Dai}, {Cheng}, {Lan}, \& {Zhang}}]{Zhong.et.al.2019}
{Zhong}, S.-Q., {Dai}, Z.-G., {Cheng}, J.-G., {Lan}, L., \& {Zhang}, H.-M. 2019, \apj, 884, 25, \dodoi{10.3847/1538-4357/ab3e48}

\end{thebibliography}

    \begin{table}[b]
        \caption{
        The Median Frequencies and Widths of QPO for Two GRBs, along with Their $\pm1\sigma$ Values of Central Frequencies
        }
        \begin{ruledtabular}
            \begin{tabular}{lcccccccc}
                \textrm{GRB} & $T_{90}(\mathrm{~sce})$ &$\nu_{1}(\mathrm{~Hz})$ & $\Delta \nu_{1}(\mathrm{~Hz})$& $\tau_{1}(\mathrm{~sce})$ & $\nu_{2}(\mathrm{~Hz})$ & $\Delta \nu_{2}(\mathrm{~Hz})$ & $\tau_{2}(\mathrm{~sce})$\\
                \colrule
                931101B & 0.034 &$877_{-8}^{+0}$ & $15_{-2}^{+7}$ & $0.067_{-0.021}^{+0.010}$& $2612_{-8}^{+9}$ & $14_{-3}^{+7}$ & $0.071_{-0.024}^{+0.019}$\\
                910711 & 0.014 & $1113_{-8}^{+7}$ & $25_{-7}^{+9}$ & $0.040_{-0.011}^{+0.016}$ & $2649_{-7}^{+6}$ & $26_{-7}^{+9}$ & $0.038_{-0.009}^{+0.015}$\\
            \end{tabular}
        \vspace{0.5em}
        \textbf{Note.} We also quote $T_{90}$ and estimate the damping times $\tau$ based on the widths for each GRB.
        \end{ruledtabular}
        \label{tab1}
    \end{table}

        \begin{table}[b]
            \caption{Overlapping Mass Ranges of Short-duration GRBs GRB 931101B and GRB 910711 for Different EOSs in Units of Solar Mass ($M_{\odot}$)}
            \begin{ruledtabular}
                \begin{tabular}{ccc}
                    \textbf{EOS} & \textrm{931101B} & \textrm{910711}\\
                    \colrule
                    \multirow{2}{*}{\textbf{GM1}} & $1.193-1.272$& $ 1.536-1.684 $ \\
                    & $1.284-1.379$ & $ 1.702-1.786 $ \\
                    \hline
                    \multirow{2}{*}{\textbf{TM1}} & $1.707-1.717$ & $ 2.034-2.080 $ \\
                    & $1.737-1.883$ & $ 2.117-2.145 $ \\
                    \hline                                            
                    \multirow{2}{*}{\textbf{NL3}} & $1.737-1.772$ & $ 2.224-2.320 $ \\
                    & $1.791-2.008$ & $ 2.435-2.521 $ \\
                    \hline                                            
                    \multirow{2}{*}{\textbf{APR}} & \multirow{2}{*}{$1.543-1.704$} & $ 1.914-1.969 $ \\
                    & & $ 2.057-2.090 $ \\
                    \hline                                            
                    \multirow{2}{*}{\textbf{SLy4}} & \multirow{2}{*}{$1.621-1.728$} & $ 1.910-1.946$  \\
                    & & $ 2.001-2.014 $ \\
                    \hline                                            
                    \textbf{DDME2}  & $2.185-2.276$ & $ 2.437-2.478 $ \\
                \end{tabular}
            \end{ruledtabular}
            \label{tab2}
        \end{table}

\bibliographystyle{aasjournal}
\end{document}